\documentclass[leqno,10pt]{amsart}

\usepackage{amsthm}
\usepackage{amsmath}
\usepackage{amssymb}
\usepackage{latexsym}
\usepackage{enumerate}
\usepackage{color}
\usepackage[dvipdfmx]{graphicx}

\newcommand{\be}{\begin{equation}}
\newcommand{\en}{\end{equation}}
\newcommand{\ea}{\end{eqnarray}}
\newcommand{\ba}{\begin{eqnarray}}
\newcommand{\ean}{\end{eqnarray*}}
\newcommand{\ban}{\begin{eqnarray*}}

\begin{document}
\title[Expected exponential utility maximization of insurers]
{Expected exponential utility maximization of insurers with a general diffusion factor model : The complete market case.
\footnote{Preliminary-version(\today).}}

\author[Hiroaki Hata]{Hiroaki Hata}
\thanks{Hiroaki Hata's research is 
supported by a Grant-in-Aid for Young Scientists (B), 
No. 15K17584, from 
Japan Society for the Promotion of Science.}
\address{Hiroaki Hata}
\address{Department of Mathematics, Faculty of Education, Shizuoka University, Ohya, Shizuoka, 422-852, Japan}
\email{hata@shizuoka.ac.jp}

\author[Shuenn-Jyi Sheu]{Shuenn-Jyi Sheu}
\address{Shuenn-Jyi Sheu}
\address{Department of Mathematics, National Central University, Zhongda Rd., Zhongli District, Taoyuan City 32001, Taiwan}
\email{sheusj@math.ncu.edu.tw}
 
\author[Li-Hsien Sun]{Li-Hsien Sun}
\address{Li-Hsien Sun}
\address{Graduate Institute of Statistics, National Central University, Zhongda Rd., Zhongli District, Taoyuan City 32001, Taiwan}
\email{lihsiensun@ncu.edu.tw}


\begin{abstract} 
In this paper, we consider the problem of optimal investment by an insurer. The insurer invests in a market consisting of a bank account and $m$ risky assets. The mean returns and volatilities of the risky assets depend nonlinearly on economic factors that are formulated as the solutions of general stochastic differential equations. The wealth of the insurer is described by a Cram\'er--Lundberg process, and the insurer preferences are exponential. Adapting a dynamic programming approach, we derive  Hamilton--Jacobi--Bellman (HJB) equation. And, we prove the unique solvability of HJB equation. In addition, the optimal strategy is also obtained using the coupled forward and backward stochastic differential equations (FBSDEs). Finally, proving the verification theorem, we construct the optimal strategy. 
\end{abstract}

\maketitle

\newtheorem{df}{\bf Definition}[section]
\newtheorem{lem}{\bf Lemma}[section]
\newtheorem{thm}{\bf Theorem}[section]
\newtheorem{cor}{Corollary}[section]
\newtheorem{prop}{\bf Proposition}[section]
\newtheorem{rem}{\bf Remark}[section]
\newcommand{\myqed}{\hbox{\rule[-2pt]{3pt}{6pt}}}

\numberwithin{equation}{section}

\section{introduction}
Recently, optimization problems of insurance companies have been studied by many works. Most of these works were solved by analyzing Hamilton--Jacobi--Bellman (HJB) equations using the dynamic programming approach. We shall introduce these studies below.

Problems of minimizing ruin probabilities has been studied by \cite{AM2009, BHLT2014, B1995, HP2000, HS2004, L2008, LTT2008, LWZ2016, S2001, S2002, TM2003}.
\cite{AM2009, BHLT2014, B1995,HP2000, HS2004} treated optimal investment problems, and 
\cite{LWZ2016, S2001, TM2003} studied optimal reinsurance problems. And,
\cite{L2008, LTT2008, S2002} considered optimal investment and reinsurance problems.

\cite{BF2013, FHMS2008, HY2017, W2007, XZY2017, YZ2005} studied optimal investment problems for maximizing expected exponential utilities. \cite{FHMS2008, W2007, YZ2005} employed Black-Scholes models. And, \cite{BF2013, BFS2018, HY2017} adopted stochastic factor models that can be expected to compensate for the disadvantages of Black-Scholes model. 

Mean-variance insurer's optimal investment-reinsurance problems have been studied by \cite{LRZ2015, LL2013, LZL2012, SZ2015, ZL2011}.
\cite{GL2014} studied optimal investment-reinsurance problem for maximizing a power utility criterion.
\cite{BG2008, HYZ2016, LB2014, LYG2011, SRZ2014, XZY2017, WRZ2018, ZMZ2016, ZRZ2013} treated optimal investment-reinsurance problems for maximizing exponential utility criterions.

Moreover, \cite{BW2017} studied an optimal investment and risk control problem.



On the other hand, \cite{WXZ2007, ZC2014, Z2009} use the martingale method 
based on equivalent martingale measures and martingale representation theorems.
Wang et al. 
\cite{WXZ2007} treated an expected exponential utility maximization with Black--Scholes model. And, Zhou \cite{Z2009} considered a counterpart of \cite{WXZ2007} with the risky asset described by a L$\acute{\rm{e}}$vy process. Furher, Zhou-Cadenillas \cite{ZC2014} studied an optimal investment and risk control problem.

In addition, the stochastic maximum principle and forward backward stochastic differential equations (FBSDEs) are the main tools to obtain the solutions, see  \cite{Pham2009,Touzi2012}. Hu et al. \cite{HU2005} studied the utility optimization problem in incomplete market through the FBSDE approach. Cheridito-Hu \cite{CH2011} analyzed the optimal consumption and investment  problem with general constraints in complete market.  Horst et al. \cite{HI2014} considered the utility optimization problem with liability. The portfolio optimization under nonlinear utility is discussed in \cite{HKT2016}. Sekine \cite{Sekine2006} discussed exponential hedging by applying BSDEs approach. Shen-Zeng \cite{SZ2015} solved mean-variance optimal investment-reinsurance problem by adapting BSDEs approach.

\color{blue}The existence and uniqueness of BSDEs with the Lipschitz generators and the squared integrable terminal condition are proved in \cite{PP1990}. Howerver, based on the feature of the Merton type problems, the corresponding FBSDEs with quadratic growth are obtained. The existence of quadratic backward stochastic differential equations (QBSDEs) with the bounded terminal condition is discussed in \cite{Ko2000,TE2008}. In particular, Bahlali et al \cite{KMY2017} and Briand and Hu \cite{BH2006,BH2008} analyzed QBSDE with unbounded terminal values and $L^2$-terminal data respectively. Barrieu and El Karoui \cite{PE2013} studied the unbounded quadratic BSDEs. Imkeller  and Reis \cite{IR2010} discussed the path regularity for BSDEs with truncated quadratic growth and Frei et al. \cite{Frei2014}, Cheridito and Nam \cite{CN2015}, Hu and Tang \cite{YS2016}, and Jammneshan et al. \cite{JKP2014} considered the multi-dimensional backward stochastic differential equations. The results of one dimensional superquadratic BSDEs are shown in \cite{CN2014,DHB2011}. 

Peng and Wu \cite{PW1999} proposed the existence and uniqueness of fully coupled FBSDEs based on the monotonicity conditions. Delarue \cite{De2002} discussed the existence and uniqueness of FBSDEs in a non-degenerate case based on the connection with the quasi-linear parabolic system of PDEs.  The well posedness of FBSDEs where the coefficients are uniformly Lipschitz are analyzed in \cite{MWZZ2015} using the decoupling random field. The solvability of coupled FBSDEs with quadratic growth is studied in \cite{LT2015,LT2017} through the Bounded Mean Oscillation (BMO)-martingale technique conditional on the small time duration. Kupper et al. \cite{KLT2017} analyzed the local and global existence and uniqueness results for multidimensional coupled FBSDEs with superquadratic growth using the Malliavin calculus and PDE technique. Note that in the previous discussion, the coupled FBSDEs do not have the jumps. The uniqueness of the coupled FBSDEs with jumps is difficult to be verified using the PDE approach due to the challenge given by the regularity for partial integro-differential equations (PIDE). In addition, in order to show the existence and uniqueness of FBSDEs using the BMO method, the jump terms is needed to be bounded See \cite{MM2010}.  This condtion is not suitable for the proposed problem since the claims for the insurer are unbounded. In this paper, we study the particular coupled FBSDEs with jumps motivated by the optimization portfolio problem for the insurer. The uniqueness of the corresponding FBSDEs with jumps can be obtained using the Jensen's inequality and martingale technique.

\color{black}
In particular, we state Badaoui-Fern$\acute{{\rm a}}$ndez \cite{BF2013}, Badaoui et al. \cite{BFS2018}, Fern$\acute{\text{a}}$ndez et al. \cite{FHMS2008}, Hata-Yasuda \cite{HY2017},  Wang \cite{W2007}, Yang-Zhang \cite{YZ2005}, Zhou \cite{Z2009}.
These treated the problems of optimal investment by insurance companies when the utility function is of exponential type.


These problems can be often solved by using dynamic programming approach.
Browne \cite{B1995}, Fern$\acute{\text{a}}$ndez et al. \cite{FHMS2008}, Wang \cite{W2007}, and Yang-Zhang \cite{YZ2005} considered Black--Scholes models.
In \cite{B1995} the risk process follows a Brownian motion with drift. In \cite{FHMS2008} the classical Cram$\acute{\text{e}}$r--Lundberg model was adopted as the risk process. In \cite{W2007} the claim process was a pure jump process. 
In \cite{YZ2005} the risk process was a compound Poisson process perturbed by a standard Brownian motion.

Badaoui-Fern$\acute{{\rm a}}$ndez \cite{BF2013} and Badaoui et al. \cite{BFS2018} considered stochastic volatility models as a counterpart of \cite{FHMS2008}.  Note that in \cite{BF2013}, the correlation between the risky asset and the factor process is zero. \cite{BFS2018} allowed that the risky assets and factor processes are correlated.
Hata-Yasuda \cite{HY2017} treated a linear Gaussian stochastic factor model, and constructed the explicit optimal strategy. 
And, \cite{LYG2011} treated the optimal investment and reinsurance problem with an Ornstein-Uhlenbeck model
Moreover, Xu et al.
\cite{XZY2017} studied the optimal investment and reinsurance problem counterpart of \cite{BF2013}.

Our objective is to extend previous work \cite{HY2017} to a more general stochastic factor model that the riskless interest rate is not constant.  
Indeed, we consider a market consisting of one bank account and $m$ risky stocks. 
Suppose that the bank account process $S^{0}$ and the price processes of the risky stocks $S:=(S^{1}, \cdots ,S^{m})^{*}$ (here, we denote $(\cdot)^{*}$ the transpose of a vector or a matrix. ) are governed by the equations:   
\begin{align}  
dS^{0}_{t}&=S^{0}_{t}r(Y_{t})dt, \quad S^{0}_{0}=s_0^{0}, \nonumber \\ 
dS^{i}_{t}&=S^{i}_{t}\left\{\mu^{i}(Y_{t})dt+\sum _{k=1}^{m}\sigma_p^{ik}(Y_{t})dW^{k}_{t}\right\}, \quad S^i_0=s^i_0, \nonumber \\
dY_{t}&=g(Y_{t})dt+\sigma_f(Y_{t})dW_{t}, \quad Y_0=y\in \mathbb{R}^{n}, \label{factor} 
\end{align}
where $(W_t)_{t\ge0}$ 
is an $m$ dimensional standard Brownian motion process defined on an underlying probability space $(\Omega ,\mathcal{F}, P)$. 
And, $\sigma_p$ and $\sigma_f $ are $m\times m$, $n\times m$ matrix-valued functions respectively and $\mu$ and $g$ are $\mathbb{R}^{m}$-valued, $\mathbb{R}^{n}$-valued functions respectively.

Consider an insurer who invests at time $t$ the amount $\pi ^{i}_{t}$ of wealth $X_t^{\pi}$ in the $i$th risky asset with price $S^{i}_t, \ i=1,\ldots, m$. With $\pi_{t}=(\pi ^{1}_{t},\ldots \pi ^{m}_{t})^{*}$ chosen, the amount of his wealth invested in the bank account is 
$$X_t^{\pi}-\pi^{*}_{t}{\bf 1}=X_t^{\pi}-\sum_{i=1}^m \pi^i_t.$$
Here ${\bf 1}=(1,\ldots ,1)^{*}$. 
Then, the insurer's wealth has the dynamics:
\begin{align}
X_t^{\pi}&=x+ct-J_t+\int^t_0 \left\{\sum^{m}_{i=1}\pi^i_s \frac{dS^{i}_s}{S^{i}_s}+\left(X^\pi_s-\pi^{*}_s{\bf 1} \right)\frac{dS^{0}_s}{S^{0}_s} \right\} \nonumber  \\
&=x+\int^t_0 \left\{c+\pi^*_s(\mu(Y_s)-r(Y_s){\bf 1})+r(Y_s)X^{\pi}_s \right\}ds+\int^t_0 \pi^*_s\sigma_p(Y_s) dW_s-J_t, \label{wealthequation}
\end{align}
where $x$ is the initial surplus and $c>0$ is the premium rate. And, the process $J_t$ is defined by
$$J_t:=\sum^{p_t}_{i=1} Z_i,$$
where $(p_t)_{t\ge0}$ is a Poisson process with a constant intensity $\lambda>0$, and $(Z_i)_{i \ge 1}$, the claim sizes, is a sequence of independent non-negative random variables with identical distribution $\nu$. 
Moreover, we assume that
\begin{align}
\int_{z>0}z \nu(dz)<\infty.
\end{align}
This gives the moment condition of claim size which will be cited later in this paper.
A stronger moment condition will be stated later (see {\bf (A6)} in Theorem \ref{Thm-HJB}).
We also assume that $(W_t)_{t\ge0}, (p_t)_{t\ge0}$ and $(Z_i)_{i \ge 1}$ are mutually independent.
Moreover, for each $t>0$ the filtration $(\mathcal{F}_t)_{t\ge0}$ is defined by
$$\mathcal{F}_t:=\sigma \{ W_s, p_s, Z_j{\bf 1}_{j\le p_s}; s\le t, j\ge 1 \}.$$
Then, we write Poisson random measure of $J$ on $[0, \infty) \times [0, \infty)$ as $N(dt, dz)$:
$$N([0, t] \times U):=\sum_{0\le s \le t} {\bf 1}_{U}(\Delta J_s)=\sum^{p_t}_{i=1} {\bf 1}_{U}(Z_i)$$
for a Borel set $U \subset [0, \infty)$, where $\Delta J_s:=J_s-J_{s-}$. We have
$$
J_t=\int_0^t \int_{z>0} z  N(ds, dz).
$$
Then, we also define the compensated Poisson random measure
$$\tilde{N}(dt, dz):=N(dt, dz)-\lambda \nu(dz)dt.$$

For simplicity, we always assume $r, \mu, g, \sigma_p$ and $\sigma_f$ are sufficiently smooth.
We also assume the following conditions: 
\begin{itemize} 
\item [{\bf (A1)}] $r, \mu, g, \sigma_p$ and $\sigma_f$ are globally Lipschitz smooth such that their first order and second order derivatives are of linear growth.
\item [{\bf (A2)}] $\sigma_p (x)$ is invertible.
\item [{\bf (A3)}] There are constants $\mu_1,\mu_2>0$ such that for $x, \eta \in \mathbb{R}^n$, $\xi\in \mathbb{R}^m$,
  \begin{align*}
&\mu _{1}|\xi |^{2}\le \xi^{*}\sigma_p (x) \sigma_p (x)^{*}\xi \le \mu _{2}|\xi |^{2},\\
&\mu _{1}|\eta |^{2}\le \eta^{*}\sigma_f (x) \sigma_f (x)^{*}\eta \le \mu _{2}|\eta |^{2}.
\end{align*} 
\item [{\bf (A4)}] $r$ satisfies
$$0 \le r(y) \le \bar{r},$$
where $\bar{r}$ is a positive constant.
\end{itemize}
\begin{rem}
The smooth of the coefficients will be needed to show the solution of HJB equation is smooth. In particular, in several places we will use Theorem $10$, Section $9$, Chapter $2$ in \cite{K1980}. See the proof of Theorem $\ref{ab thm}$, where we also need the growth condition in {\bf (A1)}.
\end{rem}

In this paper, define an expected utility of the terminal wealth : for a given constant $T>0$,  
\begin{align}\label{criterion}
J(T, x, y ;\pi):=E\left[ U(X^{\pi}_{T})\right].
\end{align} 
Here $U : \mathbb{R} \to \mathbb{R}$ is an exponential type utility function, i.e.
\begin{itemize} 
\item[{\bf (A5)}] \qquad \qquad  \qquad　$\displaystyle U(x):=-{\mathrm e}^{-\alpha x}, \ \alpha>0$. 
\end{itemize}
Uisng dynamic programming approach and FBSDE approach, we consider the following problem :
\begin{enumerate}
  \item [{\bf (P)}] \qquad \qquad  \qquad　$\displaystyle \sup_{\pi \in \mathcal{A}_{T}} J(T, x, y ;\pi)$.
\end{enumerate}
Here $\mathcal{A}_{T} (\subset \mathcal{L}_{2,T})$ is the set of admissible strategies, 
where $\mathcal{L}_{2,T}$ is defined by
\begin{align*}
{\mathcal L }_{2,T}:=\left\{(\pi_s)_{s\in[0, T]} ; \pi_s \ \text{is a} \ \mathbb{R}^m\text{-valued}\ \mathcal{F}_s\text{-progressively measurable}  \right. \\
\left. \text{stochastic process such that}  \int^T_0 |\pi_s|^2 ds<\infty, P-a.s. \right\}.
\end{align*}
The precise definition will be given later in this paper.

In Section \ref{DPA}, applying dynamic programming approach, we consider the problem {\bf (P)}. For that, we derive a HJB equation.
In Subsection \ref{EU}, the property of the complete market will help us to get the solution of the HJB equation.
In Subsection \ref{VT}, we prove the verification theorem using the HJB equation and its solution.
In Section \ref{FBSDEA}, using FBSDE approach, we consider the problem {\bf (P)}. First, using the Pontryagin maximum principle, we derive the EBSDEs.
In Subsection \ref{FBSDEAsol} we obtain the solution of FBSDEs using the property of the complete market.
One of our contributions is to show the uniqueness of the solution. Our method is analytical and technical, but it is an unusual method not seen elsewhere.
In Subsection \ref{VTFBSDEA}, using FBSDE and their solutions, we prove the verification. Our method is analytical and a good use of the nature of our setting. This is one of our contributions.
Note that the set of admissible strategy in FBSDE approach is different from that in the dynamic programming approach.
This will come from the difference between each other's approaches. Finally, we check the identity for solutions, optimal strategies and optimal values from dynamic programming approach and FBSDE approach.

\section{Dynamic programming approach}\label{DPA}
In this section, we derive Hamilton-Jacobi-Bellman (HJB) equation for {\bf (P)}.
We prepare the dynamic version :
\begin{enumerate}
  \item [{\bf (P')}] \qquad \qquad $\displaystyle V(t, x, y):=\sup_{\pi \in \mathcal{A}_{t,T}} E\left[ U(X^{\pi}_{t,T})| \mathcal{F}_t\right]$,
\end{enumerate}
where $\mathcal{A}_{t,T}$ is the restriction of the space $\mathcal{A}_T$ on the interval $[t,T]$ and $X^{\pi}_{t,T}:=X^{\pi}_{T}/X^{\pi}_{t}$.

Following a standard argument (\cite{FS2006}, Chapter IV), we can formally derive the HJB equation for {\bf (P')} with dynamics $(\ref{factor})$ and  $(\ref{wealthequation})$. The HJB equation is given by:
\begin{align} 
\begin{split}
& \displaystyle  \hspace{-0.5cm} \sup_{\pi\in \mathbb{R}^{m}}\left[\frac{\partial V}{\partial t}+\frac{1}{2}\pi^*\sigma_p(y) \sigma_p(y)^* \pi V_{xx}+\frac{1}{2}\text{tr}(\sigma_f(y) \sigma_f(y)^{*}V_{yy}) \right.\\
& \displaystyle  \left. +\pi^* \sigma_p(y) \sigma_f(y)^* V_{xy}+\{c+\pi^*(\mu(y)-r{\bf 1})+r(y)x\}V_x+g(y)^{*}V_y \right. \\
& \left. +\lambda \int_{z>0}\{V(t, x-z, y)-V(t, x, y) \}\nu(dz) \right]=0, \\
& \hspace{-0.5cm} V(T, x, y)=U(x).
\end{split} \label{HJB0}
\end{align}
Here $V_{x}, V_{y}, V_{xx}, V_{yy}, V_{xy}$ are the first order and second partial derivatives of $V$ with respect to $x, y$.
If we assume 
\begin{align}\label{ST}
V(t,x,y):=-{\mathrm e}^{-v(t,x,y)},
\end{align}
then $v$ solves the following partial differential equation :
\begin{align} 
\begin{split}
\displaystyle  \sup_{\pi \in \mathbb{R}^m}\mathcal{L}^\pi v(t, x, y)=0, \ \  v(T, x, y)=\alpha x,
\end{split} \label{HJB}
\end{align}
where $\mathcal{L}^\pi v(t, x, y)$ is defined by
\begin{align} 
\begin{split}
&\mathcal{L}^\pi v(t, x, y):= \frac{\partial v}{\partial t}+\frac{1}{2}\text{tr}(\sigma_f(y) \sigma_f(y)^{*}v_{yy})-
\frac{1}{2}v_y^{*}\sigma_f(y) \sigma_f(y)^{*}v_y +g(y)^{*}v_y\\
& +\{c+r(y)x \}v_x +\frac{v_{xx}-(v_x)^2}{2}\pi^{*}\sigma_p(y) \sigma_p(y)^{*}\pi+\pi^{*}\{(\mu(y)-r(y){\bf 1})v_x  \\
&+\sigma_p(y) \sigma_f(y)^{*}(v_{xy}-v_xv_y)\} -\lambda \int_{z>0}\left({\mathrm e}^{-v(t, x-z, y)+v(t,x,y)}-1 \right)\nu(dz).
\end{split} \label{mathcalL}
\end{align}
If 
$$v_{xx}-(v_x)^2<0$$
holds, we see that the maximizer is 
\begin{align*}
\check{\pi}(t, x, y)&:=-(\sigma_p(y)^*)^{-1}\frac{\{\theta(y)-\sigma_f(y)^* v_y\} v_x+\sigma_f(y)^* v_{xy}  }{v_{xx}-(v_x)^2}.
\end{align*}
Set
\begin{align}\label{theta}
\theta(y):=\sigma_p(y)^{-1}(\mu(y)-r(y){\bf 1}).
\end{align}
Then, we rewrite $(\ref{HJB})$ as
\begin{align} 
\begin{split}
&\frac{\partial v}{\partial t}+\frac{1}{2}\text{tr}(\sigma_f(y) \sigma_f(y)^{*}v_{yy})-\frac{1}{2}v_y^{*}\sigma_f(y) \sigma_f(y)^{*}v_y +g(y)^{*}v_y \\
&+\{c+r(y)x \}v_x -\frac{1}{2\{v_{xx}-(v_x)^2 \}} \left| (\theta(y)-\sigma_f(y)^* v_y) v_x+ \sigma_f(y)^* v_{xy} \right|^2   \\
.& -\lambda \int_{z>0} \left({\mathrm e}^{-v(t, x-z, y)+v(t,x,y)}-1 \right)\nu(dz)=0,  \\
&v(T,x,y)=\alpha x.
\end{split}\label{HJBv}
\end{align}
\begin{rem}
If the riskless interest rate is constant as in \cite{BF2013, HY2017}, we use suitable translations. Then, we 
rewrite HJB equations as parabolic partial integro-differential equations which do not depend on $x$. We can solve the equation.
However, if the riskless interest rate is not constant, this approach does not work to obtain a solution of HJB equation.
In this paper, since we treat the complete market model, we are able to obtain the solution by a different approach. See Section $\ref{EU}$. However, the new approach will fail if the market is not complete. It is still a challenge to solve the HJB equation.
\end{rem}

\subsection{The solution of HJB equation}\label{EU}
By direct calculations, we have the following.
\begin{thm}\label{Thm-HJB}
Assume {\bf (A1) $\sim$ (A5)}.  Assume also 
\begin{itemize} 
\item[{\bf (A6)}] \qquad \qquad \qquad $\displaystyle \int_{z>0} {\mathrm e}^{\alpha{\mathrm e}^{\bar{r}T} z}\nu(dz)<\infty.$ 
\end{itemize}
Then, 
\begin{align} 
\bar{v}(t,x,y):=a(t,y)x+b(t,y)
\end{align}
is a solution of $(\ref{HJBv})$. Here, $a(t,y)$ and $b(t,y)$ solve
\begin{align}\label{eqa}
\begin{split}
\frac{\partial a}{\partial t}&+\frac{1}{2}{\rm tr}(\sigma_f(y) \sigma_f(y)^{*}D^2a)+\left\{g(y)-\sigma_f(y) \theta(y)\right\}^*Da \\
&-\frac{1}{a}(Da)^* \sigma_f(y) \sigma_f(y)^{*} Da+r(y)a=0,\\
a(T&,y)=\alpha,
\end{split}
\end{align} 
and
\begin{align}\label{eqb}
\begin{split}
\frac{\partial b}{\partial t}&+\frac{1}{2}{\rm tr}(\sigma_f(y) \sigma_f(y)^{*}D^2b)+\left\{g(y)-\sigma_f(y) \theta(y)  \right. \\
& \left.  -\sigma_f(y) \sigma_f(y)^{*}\frac{Da}{a} \right\}^*Db +\frac{1}{2}  \left|\theta(y)+\sigma_f(y)^* \frac{Da}{a} \right|^2 \\
& +c a - \lambda \int_{z>0} \left( {\mathrm e}^{a(t, y) z}-1 \right)\nu(dz)=0,\\
b(T&,y)=0.
\end{split}
\end{align} 
\end{thm}
Here and in the rest, $Df=(f_{y_1}, f_{y_2}, \cdots, f_{y_n})$ is the gradient of $f(y)$.
\begin{thm}\label{ab thm} Assume {\bf (A1) $\sim$ (A6)}. Then, we have the following.

\noindent{\bf 1}. $(\ref{eqa})$ has a solution : 
\begin{align}
a(t,y)=\alpha  \widehat{E}_{t,y}\left[{\mathrm e}^{-\int^{T}_t r(\widehat{Y}_s)ds } \right]^{-1}, \label{a} 
\end{align} 
where $\widehat{E}[\cdot]$ denotes the expectation with respect to the probability measure $\widehat{P}$ defined by
\begin{align}
\frac{d\widehat{P}}{dP}\bigg |_{{\mathcal{F}}_T}&=\widehat{\mathcal{E}}_T,
\end{align}
where 
\begin{align}\label{widehatmathcalE}
\widehat{\mathcal{E}}_t :=\exp\bigg(-\int_0^t\theta(Y_s)^*dW_s-\frac{1}{2}\int_0^t|\theta(Y_s)|^2ds\bigg).
\end{align}
And, $\widehat{Y}_s$ is the solution of 
\begin{align}\label{widehatY}
\begin{split}
d\widehat{Y}_s&=\left\{g(\widehat{Y}_s)-\sigma_f(\widehat{Y}_s) \theta(\widehat{Y}_s)\right\}ds+\sigma_f(\widehat{Y}_s)d\widehat{W}_s, \ s \ge t, \\
\widehat{Y}_t&=y,
\end{split}
\end{align} 
where $\widehat{W}_s$ is a Brownian motion under $\widehat{P}$ : 
\begin{align*}
\widehat W_s&=W_s+\int_t^s\theta(Y_u)du.
\end{align*} 

\noindent{\bf 2}. $(\ref{eqb})$ has a solution : 
\begin{align}
b(t,y)=& \bar{E}_{t,y}\left[\int^T_t \left\{\frac{1}{2}  \left|\theta(\bar{Y}_s)+\sigma_f(\bar{Y}_s)^* Da(s, \bar{Y}_s) a(s, \bar{Y}_s)^{-1}  \right| ^2 \right. \right. \label{b} \\
& \left. \left. +c a(s, \bar{Y}_s)-\lambda \int_{z>0} \left( {\mathrm e}^{a(s, \bar{Y}_s) z}-1 \right)\nu(dz)  \right\}ds \right],   \nonumber
\end{align}
where $\bar{E}[\cdot]$ denotes the expectation with respect to the probability measure $\bar{P}$ defined by
\begin{align}
\frac{d\bar{P}}{dP}\bigg |_{{\mathcal{F}}_T}&=\bar{\mathcal{E}}_T,
\end{align}
where 
\begin{align}\label{barmathcalE}
\bar{\mathcal{E}}_t :=&\exp\bigg(-\int_0^t\left\{\theta(Y_s)+\sigma_f(\bar{Y}_s)^* Da(s, \bar{Y}_s) a(s, \bar{Y}_s)^{-1}  \right\}^*dW_t \\ 
& -\frac{1}{2}\int_0^t\left|\theta(Y_s)+\sigma_f(\bar{Y}_s)^* Da(s, \bar{Y}_s) a(s, \bar{Y}_s)^{-1}  \right|^2ds\bigg). \nonumber
\end{align}
And, $\bar{Y}_s$ is the solution of 
\begin{align}\label{barY}
\begin{split}
d\overline{Y}_s=&\left\{g(\bar{Y}_s)-\sigma_f(\bar{Y}_s) \theta(\bar{Y}_s) -  \sigma_f(\bar{Y}_s) \sigma_f(\bar{Y}_s)^* Da(s, \bar{Y}_s) a(s, \bar{Y}_s)^{-1} \right\}ds\\
& +\sigma_f(\bar{Y}_s)d\bar{W}_s, \ s \ge t, \quad \bar{Y}_t=y,
\end{split}
\end{align} 
where $\bar{W}_t$ is a Brownian motion under $\bar{P}$ : 
\begin{align*}
\bar W_t&=W_t+\int_0^t \left\{ \theta(\bar{Y}_s)+\sigma_f(\bar{Y}_s)^* Da(s, \bar{Y}_s) a(s, \bar{Y}_s)^{-1}  \right\}ds.
\end{align*} 
\end{thm}

Before showing the proof of this theorem, we prepare the following lemma.
\begin{lem}\label{esponential-mart} $($Lemma $4.1.1$ of \cite{Bens1992}$)$
Use $(\ref{factor})$. For a given continuous function $H(t,y)$ satisfying $|H(t,y)|\le C(1+|y|)$ we define
\begin{align*}
\rho_t :=&\exp\bigg(-\int_0^t H(s, Y_s)^*dW_s -\frac{1}{2}\int_0^t\left|H(s, Y_s) \right|^2ds\bigg). \nonumber
\end{align*}
Then, we have
$$E[\rho_t]=1\quad {\rm for } \ t \in [0,T].$$
\end{lem} 

\noindent{\bf Proof of Theorem \ref{ab thm}.}
Setting
$$\hat{a}(t,y)=a(t,y)^{-1},$$
then $\hat{a}$ solves
\begin{align}\label{eqahat}
\begin{split}
\frac{\partial \hat{a}}{\partial t}&+\frac{1}{2}\text{tr}(\sigma_f(y) \sigma_f(y)^{*}D^2\hat{a})+\left\{g(y)-\sigma_f(y) \theta(y) \right\}^*D\hat{a}-r(y)\hat{a}=0,\\ 
\hat{a}&(T,y)=\alpha^{-1}.
\end{split}
\end{align} 
From Lemma \ref{esponential-mart} we see that $\widehat{P}$ is a probability measure.
Hence, using Theorem 10, Section 9, Chapter 2 in \cite{K1980}, we have
\begin{align}\label{ahat}
\hat{a}(t,y)=\alpha^{-1}  \widehat{E}_{t,y}\left[{\mathrm e}^{-\int^{T}_t r(\widehat{Y}_s)ds } \right].  
\end{align} 
Therefore, $(\ref{a})$ is obtained.

On the other hand, from ${\bf (A4)}$ we have
$$  0<a(t,y)\le \alpha {\mathrm e}^{\bar{r}T}.$$
And, from Lemma \ref{alemma} below, there is $C>0$ such that
\begin{align}\label{aproperty}
|Da(t,y)|a(t,y)^{-1} \le C(1+|y|). 
\end{align}
Hence, using Lemma \ref{esponential-mart} again, we see that $\bar{P}$ is well-defined. And, we can check the solvability of (\ref{barY}). 
Using Theorem 10, Section 9, Chapter 2 in \cite{K1980} again, we obtain $(\ref{b})$.

\begin{lem}\label{alemma}
Assume {\bf (A1) $\sim$ (A5)}.  Then, there are $C_1>0$ and $C_2>0$ such that $(\ref{aproperty})$ is satisfied.
\end{lem}
\begin{proof}
Setting $\varphi(t,y):=\log a(t,y)$, we have 
\begin{align}\label{eqphi}
\begin{split}
&\frac{\partial \varphi}{\partial t}+\frac{1}{2}\text{tr}(\sigma_f(y) \sigma_f(y)^{*}D^2\varphi)-\frac{1}{2}(D\varphi)^*\sigma_f(y) \sigma_f(y)^{*}D\varphi \\
&\qquad +\left\{g(y)-\sigma_f(y) \theta(y) \right\}^*D\varphi+r(y)=0,\\
&\varphi(T,y)=\log \alpha.
\end{split}
\end{align}
In the rest of the proof, we may use the result in Krylov \cite{K1980} (Theorem 10, Section 9 in Chapter 2).
Noting that $\frac{\partial a}{\partial t}<0$ holds, we see that $\frac{\partial \varphi}{\partial t}<0$ holds.
Using $\frac{\partial \varphi}{\partial t}<0$ and following the arguments of Lemma 3.5 of \cite{HNS2017} or Theorem 2.1 of \cite{N2003}, we see that 
\begin{align}
\begin{split}
&|D \varphi(t, y)|^2-\kappa \frac{\partial  \varphi}{\partial t}(t, y) \le  C'\left\{|D (\sigma_f\sigma_f^*)|^2_{2r}+|\sigma_f\sigma_f^*|^2_{2r} \right.  \label{HJB gradient estimate} \\
& \left.  +|g-\sigma_f\theta|^2_{2r}+|D (g-\sigma_f\theta)|_{2r}+|r|_{2r}+|D r|_{2r}+1 \right\}, \ y\in B_r, t\in [0, T], 
\end{split}
\end{align}
where $C'$ is a positive constant independent of $r$ and $T$, $\kappa$ is sufficiently large constant, and $|\cdot|_{2r}:=||\cdot||_{L^{\infty}(B_{2r})}$. Hence, there is $C>0$ such that 
$|D\varphi(t,y)|\le C(1+|y|)$ on $[0,T] \times \mathbb{R}^n.$
\end{proof}

\subsection{Verification theorem}\label{VT}
In this section, we show the verification theorem for {\bf (P)}. 
Define the set of admissible strategies defined by
\begin{align*}
\mathcal{A}_{T}:=\left\{ (\pi_t)_{t\in [0,T]} \in \mathcal{L}_{2,T} ;   E[\mathcal{E}_{T}(\pi)]=1\right\},
\end{align*}
where $\mathcal{E}_{t}(\pi)$ is defined by
\begin{align*}
&\mathcal{E}_{t}(\pi):= \exp \left(-\int^t_0 h(s, X^{\pi}_s, Y_s, \pi_s)^* dW_s -\frac{1}{2} \int^t_0 |h(s, X^{\pi}_s, Y_s, \pi_s)|^2ds\right) \\
&\cdot \exp \left( \int^t_0 \int_{z>0} a(s, Y_{s}) z N(ds, dz)+\lambda \int^t_0 \int_{z>0}\left(1-{\mathrm e}^{a(s, Y_{s}) z} \right)\nu(dz) ds \right).
\end{align*}
Here, $h(t,x,y, \pi)$ is defined by
\begin{align*}
h(t,x,y, \pi):=\sigma_f(y)^*\{Da(t,y)x+Db(t,y)\}+\sigma_p(y)^*\pi a(t,y).
\end{align*}

Then, we have the following.
\begin{thm}\label{verification theorem}
Assume {\bf (A1) $\sim$ (A6)}. Assume also 
\begin{itemize} 
\item[{\bf (A7)}] \qquad \qquad \qquad $\displaystyle \int_{z>0} {\mathrm e}^{\alpha{\mathrm e}^{2 \bar{r}T} z}\nu(dz)<\infty.$ 
\end{itemize}
Define
\begin{align}\label{op}
\begin{split}
\tilde{\pi}(t, x, y):=&a(t,y)^{-1}(\sigma_p(y)^*)^{-1}\left[\sigma_f(y)^* Da(t,y) \left(-x+a(t,y)^{-1} \right)  \right. \\
& \left. +\theta(y)-\sigma_f(y)^*Db(t,y)  \right].
\end{split}
\end{align}
Then, the strategy $\tilde{\pi}_t \in \mathcal{A}_{T}$ defined by 
\begin{align}\label{optimalstrategy}
\tilde{\pi}_t:=\tilde{\pi}(t, X^{\tilde{\pi}}_t, Y_t) 
\end{align}
is optimal for {\bf (P)}. Moreover, we have
\begin{align}\label{OV}
V(0,x,y)=\tilde{V}(0,x,y),
\end{align}
where $\tilde{V}(t,x,y)$ is defined by
\begin{align}\label{optimalvalue}
\tilde{V}(t,x,y):=-{\mathrm e}^{-a(t,y)x-b(t,y)}.
\end{align}
Here, $a(t,y)$ and $b(t,y)$ are defined by $(\ref{a})$ and $(\ref{b})$ respectively.
\end{thm}
\begin{proof}
We write $\tilde{v}(t, x, y)=a(t, y)x + b(t, y)$.
For any $\pi \in \mathcal{A}_{T}$, using $(\ref{factor})$ and  $(\ref{wealthequation})$, we have
\begin{align*}
d\tilde{v}(t, X^\pi_t, Y_t)=&\left[\tilde{\mathcal{L}}^\pi \tilde{v}(t, X^\pi_t, Y_t)dt+h(t, X^\pi_t, Y_t, \pi_t)^*dW_t-a(t, Y_{t}) \int_{z>0} z N(dt, dz)\right],
\end{align*}
where $\tilde{\mathcal{L}}^\pi v(t, x, y)$ is defined by
\begin{align*}
\tilde{\mathcal{L}}^{\pi} \tilde{v}(t, x, y)
=&\frac{\partial \tilde{v}}{\partial t}+ \frac{1}{2}{\rm tr} ( \sigma_f(y) \sigma_f(y)^* v_{yy})  + g(y)^* v_y + \{c+ (\mu(y)-r(y){\bf 1})^* \pi\} \tilde{v}_x \\
& + \frac{1}{2}\pi^* \sigma_p(y)\sigma_p(y)^*\pi \tilde{v}_{xx} + \pi^* \sigma_p(y) \sigma_f(y)^* \tilde{v}_{xy}.
\end{align*}
This can be rewritten as 
$$
d\{ -\tilde{v}(t, X^{\pi}_t, Y_t) \}
= -\mathcal{L}^{\pi}\tilde{v}(t, X^{\pi}_t, Y_t) dt +d \log {\mathcal E}_t(\pi).
$$
where $\mathcal{L}^\pi \tilde{v}(t, x, y)$ is given in $(\ref{mathcalL})$.
Hence, using the definition of $\mathcal{A}_T$ and the fact that $\mathcal{L}^\pi \tilde{v}(t, x, y)\le 0$ holds for $(t, x, y) \in [0,T]\times \mathbb{R} \times \mathbb{R}^n$, we have
\begin{align}
\begin{split}
J(T,x,y;\pi)&=\tilde{V}(0, x, y)  E\left[{\mathrm e}^{-\int^T_0\mathcal{L}^\pi \tilde{v}(t, X^\pi_t, Y_t)dt }  \mathcal{E}_{T}(\pi)\right] \\
&\le \tilde{V}(0, x, y)  E\left[\mathcal{E}_{T}(\pi)\right] \\
&=\tilde{V}(0, x, y).
\end{split}
\end{align}

Take $\pi=\tilde{\pi}$. Then, in Appendix \ref{tildepiA AP} we observe that 
\begin{align}\label{tildepiA}
 E\left[\mathcal{E}_{T}(\tilde{\pi})\right]=1.
\end{align}
Namely, $\tilde{\pi} \in \mathcal{A}_T$. Note that $\mathcal{L}^{\tilde{\pi}} \tilde{v}(t, x, y)= 0$ holds for $(t, x, y) \in [0,T]\times \mathbb{R} \times \mathbb{R}^n$. Then, we have
\begin{align}
\begin{split}
J(T,x,y;\tilde{\pi})&=\tilde{V}(0, x, y)  E\left[{\mathrm e}^{-\int^T_0\mathcal{L}^{\tilde{\pi}} \tilde{v}(t, X^{\tilde{\pi}}_t, Y_t)dt }  \mathcal{E}_{T}(\tilde{\pi})\right] \\
&= \tilde{V}(0, x, y)  E\left[\mathcal{E}_{T}(\tilde{\pi})\right] \\
&= \tilde{V}(0, x, y).
\end{split}
\end{align}
\end{proof}

\section{FBSDE approach}\label{FBSDEA}
In this section, we study the optimal strategy using the coupled FBSDEs based on the wealth process written as
\begin{align}\label{wealth-multi}
\begin{split}
X_t^{\pi}&=x+\int^t_0 \left\{r(Y_s)X^{\pi}_s +\pi^*_s(\mu(Y_s)-r(Y_s){\bf 1})+c\right\}ds\\
&\quad +\int^t_0 \pi^*_s\sigma_p(Y_s) dW_s -\int_0^t\int_{z>0} z  N(dz,ds), 
\end{split}
\end{align}
where the factor $Y$ is given by \eqref{factor}.  Recall our problem, namely
\begin{enumerate}
  \item [$(\tilde{{\bf P}})$] \qquad \qquad $\displaystyle  \sup_{\pi \in \mathcal{\tilde A}_{T}} J(T, x, y ;\pi)$,
\end{enumerate}  
where $J$ is given by \eqref{criterion}.  Here $\tilde{\mathcal{A}}_{T} (\subset \mathcal{L}_{2,T})$ is the set of admissible strategies described later.

We further assume $r$, $\mu$, $b$, $\sigma_p$, and $\sigma_f$ satisfying the conditions  {\bf (A1) $\sim$ (A6)}. The related optimization portfolio problems using the FBSDE approach are studied in \cite{HI2014,HU2005}. 

Following the ideas of \cite{HI2014}, given $\tilde x= U(x)=-{\mathrm e}^{-\alpha x}$, applying It\^o formula to $\tilde X_t^{\pi}=U(X_t^{\pi})$, and using
\begin{align*}
& U^{-1}(\tilde x)=-\frac{1}{\alpha}\log(-\tilde {x}),\quad U(U^{-1}(\tilde x))=\tilde x, \\
& \partial_xU((U^{-1}(\tilde x))=-\alpha\tilde x, \quad \partial_{xx}U((U^{-1}(\tilde x))=\alpha^2\tilde x. 
\end{align*}
and 
\begin{align}
\begin{split}
 d\tilde X^{\pi}_t=&\partial_xU(U^{-1}(\tilde X^{\pi}_t))\left(r(Y_t)U^{-1}(\tilde X^{\pi}_t)+\pi^*_t(\mu(Y_t)-r(Y_t){\bf{1}})+c\right)dt\\
  &+\frac{1}{2}\partial_{xx}U(U^{-1}(\tilde X^{\pi}_t))\pi_t^*\sigma_p(Y_t)\sigma^*_p(Y_t)\pi_tdt\\
 &+\partial_xU(U^{-1}(\tilde X^{\pi}_t))\pi^*_t\sigma_p(Y_t)dW_t+\tilde X^{\pi}_{t^-}\int_{z>0}({\mathrm e}^{\alpha z}-1)  N(dt,dz) \\
 =&-\alpha\tilde X^{\pi}_t\left(-\frac{r(Y_t)}{\alpha}\log (-\tilde X^{\pi}_t)+\pi_t^*(\mu(Y_t)-r(Y_t){\bf 1})+c\right)dt\\
 &+\frac{1}{2}\alpha^2\tilde X^{\pi}_t\pi_t^*\sigma_p(Y_t)\sigma_p^*(Y_t)\pi_tdt+\tilde X^{\pi}_t\int_{z>0}({\mathrm e}^{\alpha z}-1) \lambda \nu(dz)dt\\
&-\alpha\tilde X^{\pi}_t\pi^*_t\sigma_p(Y_t)dW_t+\tilde X^{\pi}_{t^-}\int_{z>0}({\mathrm e}^{\alpha z}-1) \tilde N(dt,dz) ,\; \tilde X_0= U(x). \label{tildex}
\end{split}
\end{align}
Here, $\tilde{N}(dt,dz)=N(dt,dz)-\lambda \nu(dz)dt$.
The Pontryagin maximum principle leads to the corresponding Hamiltonian written as 
\begin{align}
\begin{split}
  H&(t, \pi,\tilde x, y ,p,q^0,q):=\bigg(-\alpha\tilde x (-\frac{r}{\alpha}\log(-\tilde {x})+\pi^*(\mu-r{\bf{1}})+c) \\
&+\frac{1}{2} \alpha^2\tilde x \pi^*\sigma_p\sigma_p^*\pi +\tilde x\int_{z>0}({\mathrm e}^{\alpha z}-1)\lambda\nu(dz)  \bigg)p- \tilde x\left(\alpha\pi^*\sigma_pq \right) \\
&+\tilde x\int_{z>0}({\mathrm e}^{\alpha z}-1)q^0(z)\lambda\nu(dz),\label{Hamiltonian_multi}
\end{split}
\end{align}
where we omit the dependence of $r, \mu, \sigma_p, \sigma_f$ on $y$ in this display.
The first order condition $$\frac{\partial H}{\partial \pi}(\hat\pi)=0$$ gives
\be
\hat\pi=\frac{1}{\alpha}(\sigma_p \sigma_p^*)^{-1}\sigma_p \left(\theta+\frac{q}{p}\right).   \nonumber
\en
Denote 
\be
\hat\pi_t=\frac{1}{\alpha}(\sigma_p(Y_t)\sigma_p(Y_t)^*)^{-1}\sigma_p(Y_t)\left(\theta(Y_t)+\frac{q_t}{p_t}\right), \nonumber
\en
by the abuse of the notations. In the following, we also use the notations $\tilde X_t=\tilde X_t^{\hat\pi}$ and $  X_t= X_t^{\hat\pi}$. 
Here we do not assume that $\sigma_p$ is invertible at this moment. If $\sigma_p$ is invertible, then the market is complete. We will have
\be\label{optimal-FBSDE}
\hat\pi=\frac{1}{\alpha}(\sigma_p^*)^{-1}\left(\theta+\frac{q}{p}\right),
\en
and
\be \label{optimal-FBSDE2}
\hat\pi_t=\frac{1}{\alpha}(\sigma_p(Y_t)^*)^{-1}\left(\theta(Y_t)+\frac{q_t}{p_t}\right),
\en
Then \eqref{tildex} becomes 
\begin{equation}\label{FSDE-sec}
d\tilde X_t=D_p H(\hat\pi_t)dt+D_qH(\hat\pi_t)dW_t+\int_{z>0}D_{q^0}H(\hat\pi;z)\tilde N(dt,dz),\quad \tilde X_0=x_0,
\end{equation}
where 
\[
D_p H(\hat\pi_t),\quad D_qH(\hat\pi_t), \quad D_{q^0}H(\hat\pi;z)
\]
are the short version of 
\[
D_p H(\hat\pi_t, \tilde X_t,p_t,q_t,q_t^0(z)),\;D_qH(\hat\pi_t, \tilde X_t,p_t,q_t,q_t^0(z)), \; D_{q^0}H(\hat\pi_t, \tilde X_t,p_t,q_t,q_t^0(z)).
\]
There is a dependence of $H$ on $Y_t$.
In addition, $p_t$ satisfies the dual equation given by 
\begin{equation}\label{BSDE-sec}
dp_t=-D_{\tilde x}H(\hat\pi_t)dt+q_t^*dW_t+\int_{z>0}q_t^0(z)\tilde N(dt,dz), \quad p_T=1,
\end{equation} 
with $D_{\tilde x}H(\hat\pi_t)$ again the short version of
\[
D_{\tilde x}H(\hat\pi_t, \tilde X_t,p_t,q_t,q_t^0(z)).
\]
Observe that \eqref{FSDE-sec} is a forward equation which poses the initial condition, \eqref{BSDE-sec} is a backward equation which poses the terminal condition, and (\ref{FSDE-sec}) and (\ref{BSDE-sec}) together is called a pair of FBSDEs (forward-backward stochastic differential equations.) A solution is given by $\tilde X_t$, $p_t$, $q_t$, and $q_t^0(z)$. The processes $\tilde X_t$, $p_t$, and $q_t$ are progressively measurable with respect to $\{{\mathcal{F}}_t\}$ in the usual sense and $q_t^0(z)$ is progressively measurable in the sense that the restriction of the function on $[0,s]\times [0,\infty)\times \Omega$ is measurable with respect to the product $\sigma$-field given by ${\mathcal{B}}_s\otimes {\mathcal{B}}([0,\infty))\otimes {\mathcal{F}}_s$. 

When we plug \eqref{optimal-FBSDE} in the equations \eqref{FSDE-sec} and \eqref{BSDE-sec}, the FBSDEs is the pair of equations with complicated nonlinearity. 
Indeed, we obtain the corresponding FBSDEs written as 
\begin{align}\label{SDE-multi-optimal-1}
\begin{split}
&d\tilde X^{\hat\pi}_t= -\tilde X^{\hat\pi}_t\bigg\{\alpha c-r(Y_t)\log(-\tilde X^{\hat\pi}_t)+\frac{1}{2}\theta(Y_t)^*\sigma_p(Y_t)^*(\sigma_p(Y_t)\sigma_p(Y_t))^{-1}\sigma_p(Y_t)\theta(Y_t)\\
&-\frac{1}{2}\frac{q_t}{p_t}^*\sigma_p(Y_t)^* (\sigma_p(Y_t)\sigma_p(Y_t)^*)^{-1} \sigma_p(Y_t)\frac{q_t}{p_t}\bigg\}dt+\tilde X^{\hat\pi}_{t^-}\int_{z>0}({\mathrm e}^{\alpha z}-1) N(dt,dz)\\
&-\tilde X^{\hat\pi}_t \left(\theta(Y_t)+\frac{ q_t}{p_t}\right)^*\sigma_p(Y_t)^* (\sigma_p(Y_t)\sigma_p(Y_t)^*)^{-1}  \sigma_p(Y_t)dW_t, 
\end{split}
\end{align}
\begin{align}\label{BSDE-multi-optimal-1}
\begin{split}
d&p_t= \bigg\{\alpha c-r(Y_t)(\log (-\tilde X^{\hat\pi}_t)+1))-\int_{z>0}({\mathrm e}^{\alpha z}-1)\lambda\nu(dz) \\
&+\frac{1}{2}\left( \theta(Y_t)+\frac{q_t}{p_t}\right)^*\sigma_p(Y_t)^* (\sigma_p(Y_t)\sigma_p(Y_t)^*)^{-1} \sigma_p(Y_t)\left( \theta(Y_t)+\frac{q_t}{p_t}\right) \bigg\}p_tdt\\
&-\int_{z>0}({\mathrm e}^{\alpha z}-1)q^0_t(z)\lambda\nu(dz)dt+q_t^*dW_t+\int_{z>0}q_t^0(z)\tilde N(dt,dz),
\end{split}
\end{align}
with $\tilde X_0=U(x)$ and $p_T=1$.
Choosing $\tilde q^0=\frac{q^0}{p}$ and $\tilde q=\frac{q}{p}$ and using $x=-\frac{1}{\alpha}\log(-\tilde {x})$ and $\tilde p=\log p$, we have 
\begin{align}\label{FSDE-multi0}
d&X_t^{\hat\pi}=\bigg\{r(Y_t)X^{\hat\pi}_t+c +\frac{1}{\alpha} \left( \theta(Y_t)+\tilde{q}_t\right)^*\sigma_p(Y_t)^* (\sigma_p(Y_t)\sigma_p(Y_t)^*)^{-1} \sigma_p(Y_t) \theta(Y_t)   \bigg\}dt \\
&+\frac{1}{\alpha}\left(\theta(Y_t)+\tilde q_t\right)^*\sigma_p(Y_t)^*(\sigma_p(Y_t)\sigma_p(Y_t)^*)^{-1}  \sigma_p(Y_t) dW_t- \int_{z>0} z N(dt,dz), \nonumber\\
 \label{BSDE-multi0}
d&\tilde p_t=\bigg\{\alpha c+r(Y_t)(\alpha X^{\hat\pi}_t-1)  \\
&+\frac{1}{2}\left(\theta(Y_t)+\tilde q_t\right)^*\sigma_p(Y_t)^*(\sigma_p(Y_t)\sigma_p(Y_t)^*)^{-1}  \sigma_p(Y_t) \left(\theta(Y_t)+\tilde q_t\right)  -\frac{1}{2}|\tilde q_t|^2 \bigg\}dt \nonumber\\
\nonumber& -\lambda\int_{z>0} \{ {\mathrm e}^{\alpha z}(\tilde q^0_t(z)+1)-1\}\nu(dz)dt+\tilde q_t^*dW_t+\int_{z>0}\log(1+\tilde q_{t}^0(z))  N(dt,dz),
\end{align}
with the initial condition $X_0=x$ and the terminal condition $\tilde p_T=0$.

In the case of complete markets, that is, the number of risky assets and the number of Brownian motions are the same and $\sigma_p(y)$ is invertible for any $y$. Then 
$$\sigma_p(y)^* (\sigma_p(y) \sigma_p(y)^* )^{-1} \sigma_p(y)=I.$$ 
The equations can be simplified as follows. 
\begin{align}
 \nonumber d&X^{\hat\pi}_t=\bigg\{r(Y_t)X^{\hat\pi}_t+c+\frac{1}{\alpha}\bigg(|\theta(Y_t)|^2+ \theta(Y_t)^*\tilde q_t\bigg)\bigg\}dt\\
 & +\frac{1}{\alpha}\left(\theta(Y_t)+\tilde q_t\right)^* dW_t-\int_{z>0} z   N(dt,dz),\label{FSDE-multi}\\
\nonumber d& \tilde p_t=\bigg\{\alpha c+r(Y_t)(\alpha X^{\hat\pi}_t-1)+\frac{1}{2}|\theta(Y_t)|^2+\theta(Y_t)^*\tilde q_t\bigg\}dt \\
 & -\lambda \int_{z>0} \{ {\mathrm e}^{h_t(z)+\alpha z}-1\}\nu(dz)dt+\tilde q_t^*dW_t+\int_{z>0}h_t(z)  N(dt,dz),\label{BSDE-multi} 
\end{align}
with the initial condition $X_0=x$ and $\tilde p_T=0$. Here, $h_t(z):=\log(1+\tilde q_{t}^0(z))$.

\subsection{The solution of FBSDE}\label{FBSDEAsol}
In this subsection, we shall obtain the solution of FBSDEs \eqref{FSDE-multi}-\eqref{BSDE-multi}. 
Now, we define the following spaces :
\begin{align*}
{\mathbb S}^{p}:=\left\{(X_s)_{s\in[0, T]} ; X_s \ \text{is a} \ \mathbb{R}\text{-valued}\ \mathcal{F}_s\text{-progressively measurable}  \right. \\
\left. \text{stochastic process such that} \ E\left[\sup_{t\in[0,T]}|X_s|^2 \right] <\infty. \right\}, \\
{\mathbb L}^{p}:=\left\{(X_s)_{s\in[0, T]} ; X_s \ \text{is a} \ \mathbb{R}^m\text{-valued}\ \mathcal{F}_s\text{-progressively measurable}  \right. \\
\left. \text{stochastic process such that} \ E\left[\int^T_0 |X_s|^p ds \right] <\infty. \right\}, \\
{\mathbb L}_N^{p}:=\left\{(X_s)_{s\in[0, T]} ; X_s : [0,T]\times \mathbb{R}_+ \to \mathbb{R} \ \text{is a} \ \mathcal{F}_s\text{-progressively measurable}  \right. \\
\left. \text{stochastic process such that} \ E\left[\int^T_0 \int_{z>0} |X_s|^p \nu(dz) ds \right] <\infty. \right\}.
\end{align*}

Then, we have the following lemma for the solution of the coupled FBSDEs.
\begin{lem}\label{lemma-FBSDE}
Define $(X^{\hat\pi}_t, \tilde p_t, \tilde q_t, h_t(z))$ as 
\begin{align}\label{opX}
  X_t^{\hat{\pi}}:=&\frac{1}{\alpha-\eta(t,Y_t)}\left\{(\alpha-\eta(0,y))x_0+\phi(t,Y_t)-\phi(0,y)\right\}\\
 &+\frac{1}{\alpha-\eta(t,Y_t)}\bigg[\int_0^tr(Y_s)ds+\int_0^t\theta(Y_s)^*dW_s+\int_0^t\frac{1}{2}\left|\theta(Y_s)\right|^2ds \nonumber \\
 &-\int_0^t\int_{z>0} \left\{\alpha-\eta(s,Y_{s}) \right\} z   N(ds, dz)\nonumber\\
&+\lambda \int_0^t \int_{z>0} \left({\mathrm e}^{\{\alpha-\eta(s,Y_{s}\} z}-1 \right) \nu(dz) ds \bigg], \nonumber\\
\label{ansatz-multi}\tilde p_t &:=\eta(t,Y_t)X^{\hat\pi}_t+\phi(t,Y_t),\\
\label{q0-multi} h_t(z)&:=-\eta(t, Y_t)z, \\
 \nonumber \tilde q_t&:=\frac{\alpha}{\alpha-\eta(t,Y_t)}\bigg(X^{\hat\pi}_t\sigma_f(Y_t)^*D\eta(t,Y_t)\\
& +\sigma_f(Y_t)^*D\phi(t,Y_t)+\frac{1}{\alpha} \eta(t,Y_t)\theta(Y_t) \bigg),\label{q-multi}
\end{align}
where 
$\eta(t,y)$ and $\phi(t,y)$ must satisfy 
\begin{align}
\begin{split}
 & \frac{\partial \eta}{\partial t} +\frac{1}{2}{\rm tr}(\sigma_f(y)\sigma_f(y)^*D^2\eta )+\{g(y)-\sigma_f(y)\theta(y)\}^*D\eta\\
 &-\frac{1}{\eta -\alpha}(D\eta)^*\sigma_f(y)\sigma_f(y)^*D\eta-r(y)(\alpha-\eta) =0, \\
& \eta(T,y)=0,
\label{PDE-FBSDE-eta-sec} 
\end{split}
\end{align}
and
\begin{align}
\begin{split}
 & \frac{\partial \phi}{\partial t}+\frac{1}{2}{\rm tr}(\sigma_f(y)\sigma_f(y)^*D^2\phi ) +\left\{g(y)-\sigma_f(y)\theta(y) \right. \\
&\left. +\frac{1}{\alpha-\eta}\sigma_f(y)\sigma_f(y)^*D\eta \right\}^*D\phi+\frac{1}{\alpha-\eta}(D\eta)^*\sigma_f(y)\theta(y)\\
 &-\frac{1}{2}|\theta(y)|^2+r(y)-c(\alpha-\eta )+\int_{z>0} \left({\mathrm e}^{(\alpha- \eta)z }-1\right)\lambda\nu(dz)=0,\\
&\phi(T,y)=0.
\label{PDE-FBSDE-phi-sec} 
\end{split}
\end{align}
Then,  FBSDEs \eqref{FSDE-multi}-\eqref{BSDE-multi} has a solution $(X^{\hat\pi}_t, \tilde p_t, \tilde q_t, \tilde q^0_t) \in {\mathbb S}^{2} \times {\mathbb S}^{2} \times {\mathbb L}^{2} \times {\mathbb L}_N^{2}$ satisfying there is $C_1>0$ such that
\begin{align}\label{htz-condition}
h_t(z)\le C_1 z.
\end{align}
\end{lem}
\begin{proof}
The proof is given in Appendix \ref{proof-FBSDE1}.
\end{proof}

In Lemma \ref{lemma-FBSDE}, we have
\begin{equation}\label{BSDE-solution-sec}
p_t=\exp(\eta(t,Y_t) X_t+\phi(t,Y_t)). 
\end{equation}
We can see $\alpha-\eta(t, y)$ and $a(t, y)$ have the same equation (see (\ref{eqa})). Moreover, from the equation of $\varphi(t, y)=\log a (t, y)$ given in (\ref{eqphi}), we can see $\varphi(t, y)-\phi(t, y)$ and $b(t, y)$ have the same equation (see (\ref{eqb})), but with different terminal conditions, given by
$$
\varphi(T, y)-\phi(T, y)=\log \alpha,\ b(T, y)=0.
$$
We can conclude 
$$
\eta(t, y)=\alpha- a(t, y),\ \phi(t, y)=\log a(t, y)-\log \alpha - b(t, y).
$$
We have the following result.

\begin{thm} \label{Thm-FBSDE}
Assume {\bf (A1) $\sim$ (A6)}. 
Then, \eqref{PDE-FBSDE-eta-sec} has a solution : 
\begin{align}\label{etasol}
\eta(t,y)=\alpha-a(t,y), 
\end{align}
where $a(t,y)$ is given by $(\ref{a})$.
And, \eqref{PDE-FBSDE-phi-sec} has a solution $\phi(t, y)=\log a(t, y)-\log \alpha - b(t, y)$, which has the expression
: 
\begin{align}\label{phisol}
\begin{split}
&\phi(t,y)=\bar{E}_{t, y}\bigg[\int_t^T\bigg\{-a(s,\bar Y_s)^{-1} (Da(s,\bar Y_s))^* \sigma_f(\bar{Y}_s)\theta(\bar{Y}_s) -ca(s,\bar Y_s)\\
&+r(\bar Y_s)-\frac{1}{2}|\theta(\bar Y_s)|^2+\int_{z>0}\left({\mathrm e}^{a(s,\bar{Y}_s)z}-1\right)\lambda\nu(dz)\bigg\}ds\bigg],
\end{split}
\end{align}
where $\bar{Y}_s$ is given in $(\ref{barY})$.

\end{thm}

\begin{proof}
Letting $\tilde\eta=\frac{1}{\alpha-\eta}$, we obtain $\tilde\eta$ satisfying 
\be\label{PDE-tildeeta}
\frac{ \partial \tilde\eta}{\partial t}(t,y)+\left(g(y)^*-\theta(y)\sigma_f(y)\right)D\tilde\eta(t,y)+\frac{1}{2}{\rm tr}(\sigma_f(y)\sigma_f(y)^*D^2\tilde\eta(t,y))-r(y)\tilde\eta(t,y)=0,
 \en
with the terminal condition $\tilde\eta(T,y)=\frac{1}{\alpha}$.
Then, $(\ref{PDE-tildeeta})$ accords with $(\ref{eqahat})$. Hence, we have
\be\label{solution_eta}
\tilde\eta(t,y)=\alpha^{-1} \widehat{E}_{t,y}\left[{\mathrm e}^{-\int_t^Tr(\widehat{Y}_s)ds}\right],
\en
where $\widehat{Y}_s$ is given by $(\ref{widehatY})$.

Consider $\xi(t,y):=\log(\alpha-\eta(t,y))$. Then, observing $\frac{\partial(\alpha-\eta)}{\partial t}\le 0$ and $\frac{\partial \xi}{\partial t}\le 0$ and following the arguments 
of Lemma \ref{alemma} for the PDE of $\xi$ we see that 
\begin{align*}
\frac{|D\eta(t,y)|}{\alpha-\eta(t,y)}=|D\xi(t,y)|\le C(1+|y|).
\end{align*}
Hence, using the conditions {\bf (A1) $\sim$ (A6)} 
and Theorem 10, Section 9, Chapter 2 in \cite{K1980}, we obtain $(\ref{phisol})$.
\end{proof}

\begin{thm}\label{uniqueness}
Assume {\bf (A1) $\sim$ (A6)}. Then,  FBSDEs \eqref{FSDE-multi}-\eqref{BSDE-multi} has a unique solution of $(X^{\hat\pi}_t, \tilde p_t, \tilde q_t, h_t(z))  \in {\mathbb S}^{2} \times {\mathbb S}^{2} \times {\mathbb L}^{2} \times {\mathbb L}_N^{2}$ satisfying $\eqref{htz-condition}$.
\end{thm}
The proof is given in Appendix \ref{proof-uniqueness}.

\subsection{Verification theorem}\label{VTFBSDEA}
In this subsection, we show the verification theorem.
Define the probability measure $\hat{P}^N$ as 
\begin{align}\label{hatPN}
\frac{d\hat{P}^N}{dP}\bigg |_{{\mathcal{F}}_T}&=\hat{\mathcal{E}}^N_{T},
\end{align}
where $\hat{\mathcal{E}}^N_{t}$ is defined by
\begin{align}\label{mathcalEhatN}
\begin{split}
\hat{\mathcal{E}}^N_{t}:=&\widehat{\mathcal{E}}_{t} \cdot \exp \bigg( \int^t_0 \int_{z>0}\{ \alpha-\eta(s, Y_{s})\} z N(ds, dz) \\ 
&+\lambda \int^t_0 \int_{z>0}\left(1-{\mathrm e}^{\{\alpha-\eta(s, Y_{s})\} z} \right)\nu(dz) ds \bigg),
\end{split}
\end{align}
where $\widehat{\mathcal{E}}_t$ is given in $(\ref{widehatmathcalE})$.
Following the arguments of Appendix \ref{tildepiA AP} and Lemma \ref{esponential-mart}, we have $E[\hat{\mathcal{E}}^N_{T}]=1$. Hence, $\hat{P}^N$ is well-defined.
Note that under probability measure $\hat{P}^N$, $Y_t$ is given by
\begin{align}\label{YhatN}
dY_t&=\left\{g(Y_t)-\sigma_f(Y_t)\theta(Y_t)\right\}dt+\sigma_f(Y_t)d\hat W^N_t,
\end{align}
where 
\begin{align}\label{hatNW}
\hat W^N_t:=W_t+\int^t_0 \theta(Y_s)ds
\end{align}
 is a Brownian motion under $\hat{P}^N$.

Define the set of admissible strategies defined by
\begin{align}\label{tildemathcalA}
\tilde{\mathcal{A}}_{T}:=\bigg\{ (\pi_t)_{t\in [0,T]} \in \mathcal{L}_{2,T} ;  \hat{E}^N\bigg[ \int^T_0 | \pi_t |^2dt \bigg]<\infty. \bigg\}.
\end{align}

\begin{thm}\label{Thm-verification-FBSDE}   
Assume {\bf (A1) $\sim$ (A7)}.  
Define
\begin{align}
\begin{split}
\hat\pi&(t,x,y):=\frac{1}{\alpha-\eta(t,y)}(\sigma_p(y)^*)^{-1}\bigg( \theta(y)+\sigma_f(y)^*\{D\eta(t,y)x+D\phi(t,y) \}\bigg).
\label{optimal-FBSDE-sec}
\end{split}
\end{align}
Then, the strategy $\hat{\pi}_t \in \mathcal{\tilde A}_{T}$ defined by 
\begin{align}\label{optimalstrategy_FBSDE}
\hat{\pi}_t:=\hat{\pi}(t, X^{\hat{\pi}}_t, Y_t) 
\end{align}
is optimal for {$\tilde{\bf (P)}$}. Moreover, we have
\begin{align}\label{BSDEoptimalvalue}
\sup_{\pi \in \tilde{A}_T}E\left[U(X^\pi_T) \right]=E\left[U(X^{\hat{\pi}}_T) \right]
=-\frac{\alpha}{\alpha- \eta(0,y)}{\mathrm e}^{-\{\alpha-\eta(0,y)\}x+\phi(0,y)}.
\end{align}
\end{thm}

\begin{proof}
To prove (\ref{BSDEoptimalvalue}), we consider $\tilde{X}_t^{\pi} p_t$. We use the relation,
$$
\tilde{X}_t^{\pi} p_t=-\exp(\tilde{p}_t-\alpha X_t^{\pi}).
$$
We have 
\begin{align}\label{(0)}
dX_t^{\pi}= \left\{r(Y_t) X_t^{\pi} + \pi_t^*(\mu(Y_t)-r(Y_t) {\bf 1})+ c\right\} dt + \pi_t^{*}\sigma_p(Y_t) dW_t - \int_{z>0} z N(dt, dz),
\end{align}
and by (\ref{BSDE-multi}) and (\ref{q0-multi}),
\begin{align*}
d\tilde{p}_t= & \bigg\{\alpha c + \alpha r(Y_t) X^{\hat{\pi}}_t -r(Y_t)+ \frac 12 |\theta(Y_t)|^2 + \theta(Y_t)^* \tilde{q}_t \\
&-\lambda \int_{z>0} ({\mathrm e}^{\{ \alpha-\eta(t, Y_t)\}z}-1) \nu(dz)\bigg\} dt + \tilde{q}^*_t  dW_t - \eta(t, Y_t) \int_{z>0} z N (dt, dz).
\end{align*}
Then 
\begin{align}\label{(1)}
\begin{split}
d&(\tilde{p}_t-\alpha X_t^{\pi})=\bigg\{\alpha r(Y_t)(X_t^{\hat{\pi}}-X^{\pi}_t) -r(Y_t)-\alpha \pi^*_t(\mu(Y_t)-r(Y_t){\bf 1}) \\
&+\frac 12 |\theta(Y_t)|^2+ \theta(Y_t)^* \tilde{q}_t \bigg\}dt-\lambda \int_{z>0} ({\mathrm e}^{\{\alpha-\eta(t, Y_t)\} z}-1) \nu(dz) dt \\
&+ (\tilde{q}_t-\alpha \sigma_p(Y_t)^*\pi_t)^* dW_t +\int_{z>0} (\alpha-\eta(t, Y_t)) z N(dt, dz).
\end{split}
\end{align}
From (\ref{optimal-FBSDE2}) we have 
$$
\tilde{q}_t=-\theta(Y_t)+\alpha \sigma_P(Y_t)^* \hat{\pi}_t.
$$
From this and (\ref{theta}), we have 
\begin{align}
\begin{split}\label{(2)}
&\theta(Y_t)^* \tilde{q}_t -\alpha (\mu(Y_t)-r(Y_t){\bf 1})^* \pi_t\\
=&\theta(Y_t)^*(\tilde{q}_t-\alpha \sigma_p(Y_t)^*\pi_t)\\
=&\theta(Y_t)^*(-\theta(Y_t) + \alpha \sigma_p(Y_t)^* \hat{\pi}_t -\alpha \sigma_p(Y_t)^*\pi_t)\\
=&-|\theta(Y_t)|^2 +\alpha \theta(Y_t)^*\sigma_p(Y_t)^*(\hat{\pi}_t-\pi_t).
\end{split}
\end{align}
By (\ref{(1)}), (\ref{(2)}), we have
\begin{align}\label{(3)}
\begin{split}
d&(\tilde{p}_t-\alpha X_t^{\pi})=\bigg\{\alpha r(Y_t)(X_t^{\hat{\pi}}-X^{\pi}_t) -r(Y_t) -\frac 12 |\theta(Y_t)|^2 \\
 &+\alpha \theta(Y_t)^*\sigma_p(Y_t)^*(\hat{\pi}_t-\pi_t)\bigg\}dt+ (-\theta(Y_t)+\alpha \sigma_p(Y_t)^*(\hat{\pi}_t-\pi_t))^* dW_t  \\
& -\lambda \int_{z>0} ({\mathrm e}^{(\alpha-\eta(t, Y_t)) z}-1) \nu(dz) dt +\int_{z>0} (\alpha-\eta(t, Y_t)) z N(dt, dz).
\end{split}
\end{align}
Using (\ref{(0)}), we have
\begin{align}\label{(4)}
\begin{split}
d&(X^{\hat{\pi}}_t-X^{\pi}_t) \\
=&\{r(Y_t)(X^{\hat{\pi}}_t-X^{\pi}_t)+ \theta(Y_t)^* \sigma_p(Y_t)(\hat{\pi}_t-\pi_t) \}dt + (\hat{\pi}_t-\pi_t)^* \sigma_p(Y_t) dW_t\\
=&  r(Y_t)(X^{\hat{\pi}}_t-X^{\pi}_t) dt + (\hat{\pi}_t-\pi_t)^* \sigma_p(Y_t) d\hat{W}^N_t,
\end{split}
\end{align}
where $\hat{W}^N_t$ is given by $(\ref{hatNW})$.
From  (\ref{(3)}) and  (\ref{(4)}), we have
\begin{align*}
\tilde{p}_T&-\alpha X_T^{\pi}=  \tilde{p}_0-\alpha x + \alpha(X^{\hat{\pi}}_T-X^{\pi}_T)-\int_0^T r(Y_t) dt- \int_0^T \theta(Y_t)^* dW_t   \\
&-\frac 12 \int_0^T |\theta(Y_t)|^2 dt -\lambda \int_0^T  \int_{z>0} ({\mathrm e}^{(\alpha-\eta(t, Y_t)) z}-1) \nu(dz) dt \\
& +\int_0^T \int_{z>0} (\alpha-\eta(t, Y_t))z N(dt, dz)   \\
      = & \tilde{p}_0-\alpha x +  \alpha(X^{\hat{\pi}}_T-X^{\pi}_T)-\int_0^T r(Y_s) ds   +\log \hat{\mathcal{E}}_T^N,                      
\end{align*}
where $\hat{\mathcal{E}}_T^N$ is given in $(\ref{mathcalEhatN})$. Therefore,
$$
\tilde{X}^{\pi}_T p_T=U(x)p_0 {\mathrm e}^{\alpha(X^{\hat{\pi}}_T-X^{\pi}_T)-\int_0^T r(Y_t) dt }\cdot \hat{\mathcal{E}}_T^N.
$$
Then 
\begin{align}\label{(5)}
\begin{split}
E[\tilde{X}^{\pi}_T p_T]= & U(x)p_0E\left[ {\mathrm e}^{\alpha(X^{\hat{\pi}}_T-X^{\pi}_T)-\int_0^T r(Y_t) dt }\cdot \hat{\mathcal{E}}_T^N \right]\\
                                       =& U(x)p_0 \hat{E}^N\left[{\mathrm e}^{\alpha(X^{\hat{\pi}}_T-X^{\pi}_T)-\int_0^T r(Y_t) dt } \right],
\end{split}
\end{align}
where $\hat{E}^N[\cdot]$ is the expectation with respect to $\hat{P}^N$. 
From (\ref{(4)}), we have, under $\hat{P}^N$ 
\begin{align*}
{\mathrm e}^{-\int_0^T r(Y_t) dt} (X^{\hat{\pi}}_T-X^{\pi}_T)=\int_0^T {\mathrm e}^{-\int_0^t r(Y_s) ds} (\hat{\pi}_t-{\pi}_t)^* \sigma_p(Y_t) d\hat{W}_t^N.
\end{align*}
Using this and Jensen's inequality, we have
\begin{align}
&\frac 1{\hat{E}^N\left[{\mathrm e}^{-\int_0^T r(Y_t) dt}\right] } \hat{E}^N\left[ {\mathrm e}^{-\int_0^T r(Y_t) dt} {\mathrm e}^{\alpha (X^{\hat{\pi}}_T-X^{\pi}_T} ) \right]  \nonumber \\
\geq & \exp\left( \frac 1{\hat{E}^N \left[{\mathrm e}^{-\int_0^T r(Y_t) dt} \right] } \hat{E}^N\left[ {\mathrm e}^{-\int_0^T r(Y_t) dt} \alpha(X^{\hat{\pi}}_T-X^{\pi}_T)\right] \right)   \nonumber  \\
= & \exp\left( \frac 1{\hat{E}^N \left[{\mathrm e}^{-\int_0^T r(Y_t) dt} \right] } \hat{E}^N\left[ \int_0^T {\mathrm e}^{-\int_0^t r(Y_s) ds} (\hat{\pi}_t-{\pi}_t)^* \sigma_p(Y_t) d\hat{W}_t^N\right] \right)  \nonumber \\
=&1, \nonumber
\end{align}
if $\hat{\pi} \in \tilde{\mathcal{A}}_T$, see $(\ref{tildemathcalA})$.

Hence, using $(\ref{(5)})$, we have
\begin{align*}\label{(6)}
\begin{split}
E[\tilde{X}^{\pi}_T p_T]  =& U(x)p_0 \hat{E}^N\left[{\mathrm e}^{\alpha(X^{\hat{\pi}}_T-X^{\pi}_T)-\int_0^T r(Y_t) dt } \right] \\
\le & U(x)p_0 \hat{E}^N\left[{\mathrm e}^{-\int_0^T r(Y_t) dt } \right] \\
= & E[\tilde{X}^{\hat{\pi}}_T p_T].
\end{split}
\end{align*}
Therefore, we see that $\hat{\pi}$ is optimal if $\hat{\pi} \in \tilde{\mathcal{A}}_T$ holds.  The fact that $\hat{\pi} \in \tilde{\mathcal{A}}_T$ holds is proved in Appendix \ref{proof-verification-FBSDE}. 

We calculate $E[\tilde{X}^{\hat{\pi}}_T p_T]$.
Since we also observe
\begin{eqnarray}
\text{``$\widehat{Y}_t$ given in $(\ref{widehatY})$ under $\widehat{P}$''}
&\underset{\mathrm{law}}{=}& 
\text{``$Y_t$ given in $(\ref{YhatN})$ under $\hat{P}^N$''}, 
\nonumber
\end{eqnarray}
we see that from $(\ref{etasol})$
$$\hat{E}^N\left[{\mathrm e}^{-\int^T_0 r(Y_t)dt} \right]=\widehat{E}\left[{\mathrm e}^{-\int^T_0 r(\widehat{Y}_t)dt} \right]=\frac{\alpha}{\alpha- \eta(0,y)}.$$
Therefore, recalling $\tilde{X}^{\hat{\pi}}_0=-{\mathrm e}^{-\alpha x}$ and $p_0={\mathrm e}^{\eta(0,y) X^{\hat{\pi}}_0 +\phi(0,y)}$, we have
\begin{align*}
E\left[\tilde X^{\hat\pi}_T p_T\right]&=-{\mathrm e}^{-\{\alpha-\eta(0,y)\}x+\phi(0,y)}\tilde{E}\left[{\mathrm e}^{-\int^T_0 r(Y_t)dt} \right] \\
&= -{\mathrm e}^{-\{\alpha-\eta(0,y)\}x+\phi(0,y)}  \frac{\alpha}{\alpha- \eta(0,y)}.
\end{align*}
\end{proof}

\begin{cor}
By comparing {\em (\ref{eqa}-\ref{eqb})} and {\em (\ref{PDE-FBSDE-eta-sec}-\ref{PDE-FBSDE-phi-sec})} , we verify the identity for solutions from the PDE approach and the FBSDE approach based on the relation  
\begin{align}
\nonumber\eta(t,y)=&\alpha-a(t,y),\\
\nonumber \phi(t,y)=&\log\frac{a(t,y)}{\alpha}-b(t,y),
\end{align}
such that the optimal strategy \eqref{optimal-FBSDE-sec} and the optimal value \eqref{BSDEoptimalvalue} through the FBSDE approach are identical to the optimal strategy 
\begin{align*}
\tilde{\pi}(t, X^{\tilde{\pi}}_t, Y_t)=&a(t, Y_t)^{-1}(\sigma_p(Y_t)^*)^{-1} \\
 &\cdot \{\theta(Y_t)-\sigma_f(Y_t)^*(D a(t, Y_t) x+Db(t, Y_t)-Da(t, Y_t) a(t, Y_t)^{-1})\},
\end{align*}
and the optimal value is given
$$
V(0, x, y)=-{\mathrm e}^{-a(0, y)x - b(0, y)},
$$
using the HJB approach. 
\end{cor}

\appendix

\section{The proof of $(\ref{tildepiA})$.}\label{tildepiA AP}
Recall that
\begin{align}
h(s, X^{\tilde{\pi}}_s, Y_s, \tilde{\pi}_s)=\sigma_f(Y_s)^*Da(s, Y_s) a(s, Y_s)^{-1}+\theta(Y_s).
\end{align}
We consider
$$
\int_0^t \int_{z>0} a(s, Y_{s}) z N(ds, dz)=\sum_{i=1}^{p_t} a(T_i, Y_{T_i}) Z_i,
$$
where $Z_1, Z_2, \cdots$ are iid with distribution $\nu(\cdot)$ and 
$$
T_i=S_1+S_2+\cdots+ S_i,
$$
$S_1, S_2, \cdots$ are iid having exponential distribution with parameter $\lambda$. Then, we have
\begin{align*}
 E\left[\exp \left(\sum_{i=1}^{p_t} a(T_i, Y_{T_i}) Z_i \right)\right]=\sum_{N=0}^{\infty}  E\left[\exp \left(a(T_i, Y_{T_i}) Z_i\right); p_t=N \right].
\end{align*}

The expectation of 
$$
\exp\left(\sum_{i=1}^{N} a(T_i, Y_{T_i}) Z_i \right) {\bf 1}_{p_t=N}
$$
with respect to $S_1, S_2, \cdots$ and $Z_1, Z_2, \cdots$ while keeping $W_t, Y_t$ fixed is given by
\begin{align*}
&\int_0^t\lambda {\mathrm e}^{-\lambda t_1} \int_{z_1>0} {\mathrm e}^{a (t_1, Y_{t_1}) z_1} \nu(dz_1)  dt_1\int_{t_1}^t \lambda {\mathrm e}^{-\lambda ( t_2-t_1)} \int_{z_2>0} {\mathrm e}^{a (t_2, Y_{t_2}) z_2} \nu(dz_2)dt_2 \\
&\ \ \cdots  \int_{t_{N-1}}^t \lambda {\mathrm e}^{-\lambda (t_N-t_{N-1})}  {\mathrm e}^{-\lambda(t-t_N)} \int_{z_{N}>0} {\mathrm e}^{a (t_N, Y_{t_N}) z_N}  \nu(dz_N) dt_N\\
&=\lambda^N {\mathrm e}^{-\lambda t}  \int_0^t \int_{z_1>0} {\mathrm e}^{a (t_1, Y_{t_1}) z_1} \nu(dz_1)  dt_1\int_{t_1}^t \int_{z_2>0}  {\mathrm e}^{a (t_2, Y_{t_2}) z_2}  \nu(dz_2)dt_2 \\ 
&\ \ \cdots \int_{t_{N-1}}^t \int_{z_{N}>0}  {\mathrm e}^{a (t_N, Y_{t_N}) z_N} \nu(dz_N) dt_N\\
\displaystyle & =\frac{1}{N!} {\mathrm e}^{-\lambda t}  \left\{ \lambda \int_0^t ds \int_{z>0} \exp(a (s, Y_{s}) z) \nu(dz) \right\}^N.
\end{align*}
Therefore, the expectation of 
$$
\exp\left(\int_0^t \int_{z>0} a(s, Y_{s}) z N(ds, dz) \right)
$$
with respect to $T_1, T_2, \cdots$ and $Z_1, Z_2, \cdots$ keeping $W_s, Y_s, 0\leq s\leq t$ fixed is given by
$$
{\mathrm e}^{-\lambda t} \exp\left(\lambda \int_0^t \int_{z>0} \exp(a(s, Y_s) z)\nu(dz) ds \right).
$$
Using the expression of $\mathcal{E}_t(\tilde{\pi})$ 
and the fact that $h(s, X^{\tilde{\pi}}_s, Y_s, \tilde{\pi}_s)=\tilde{h}(s, Y_s)$ is a function of $(s, Y_s)$, we only need to work on the proof of 
\begin{align}\label{EXc}
E[\mathcal{E}^c_T]=1,
\end{align}
where $\mathcal{E}^c_t$ is defined by
\begin{align*}
\mathcal{E}^c_t:= \exp \left(-\int_0^t \tilde{h}(s, Y_s) dW_s-\frac 12 \int_0^t |\tilde{h}(s, Y_s)|^2 ds\right).
\end{align*}
Then, we observe that $|\tilde{h}(s, y)|\le C(1+|y|)$.  Using Lemma \ref{esponential-mart}, we have $(\ref{EXc})$.

\begin{rem}
The above argument may not work for general $\pi$, since $h(s, X_s^{\pi}, Y_s, \pi_s)$ depends on $N(dt, dz)$ in general. In the argument, we can not fix $\pi_s$ and take expectation with respect to $N(dt, dz)$ in the very beginning. However, if $h(s, X_s^{\pi}, Y_s, \pi_s)$ depends only on $Y_s$, then we can fix $W$ in the calculation and take expectation on $N(\cdot, \cdot)$ first as in Appendix \ref{tildepiA AP}.  

For general $\pi$, we use It$\hat{{\rm o}}$'s formula to derive
$$
d\mathcal{E}_t(\pi)= \mathcal{E}_{t-}(\pi)\left\{ -h(t, X_t^{\pi}, Y_t, \pi_t) dW_t+\int_{z>0} ({\mathrm e}^{a(t, Y_t) z}-1)\tilde{ N}(dt, dz)\right\}.
$$
Hence, it is a positive local martingale, and hence is a supermartingale and may not be a martingale unless additional conditional on $\pi$ is imposed.
\end{rem}

\section{The Proof of Theorem \ref{Thm-FBSDE}}\label{proof-FBSDE1}


Applying It\^o formula to \eqref{ansatz-multi} implies
\begin{align}
\begin{split}
  &d\tilde p_t=\bigg\{X^{\hat\pi}_t\bigg(\frac{\partial \eta}{\partial t}(t,Y_t)+\frac{1}{2}{\rm tr}(\sigma_f(Y_t)\sigma_f(Y_t)^*D^2\eta(t,Y_t))+g(Y_t)^*D\eta(t,Y_t) \\
&+r(Y_t)\eta(t,Y_t)\bigg)+\frac{\partial \phi}{\partial t}(t,Y_t)+\frac{1}{2}{\rm tr}(\sigma_f(Y_t)\sigma_f(Y_t)^*D^2\phi(t,Y_t))+g(Y_t)^*D\phi(t,Y_t)\\
&+c\eta(t,Y_t)+\frac{1}{\alpha}(\eta(t,Y_t)\theta(Y_t)+\sigma_f(Y_t)^*D\eta(t,Y_t))^* (\theta(Y_t)+\tilde q_t) \bigg\}dt\\
  &+\bigg\{X^{\hat\pi}_t(D\eta(t,Y_t))^*\sigma_f(Y_t)+(D\phi(t,Y_t))^*\sigma_f(Y_t)+\eta(t,Y_t)\frac{1}{\alpha}(\theta(Y_t)+\tilde q_t)^* \bigg\}dW_t\\
 &-\int_{z>0} \eta(t,Y_t) z  N(dt,dz). \\ 
 \label{BSDE-multi-2}
\end{split}
\end{align}
Identifying the diffusion terms in \eqref{BSDE-multi} and \eqref{BSDE-multi-2}, we obtain \eqref{q0-multi} and \eqref{q-multi}.
Inserting \eqref{q0-multi} and \eqref{q-multi} into \eqref{BSDE-multi} and \eqref{BSDE-multi-2} leads to  
\begin{align}\label{BSDE-multi-3}
\begin{split}
  &d\tilde p_t=\bigg\{X^{\hat\pi}_t\bigg(\alpha r(Y_t)+\frac{\alpha}{\alpha-\eta(t,Y_t)}\theta(Y_t)^*\sigma_f(Y_t)^*D\eta(t,Y_t)\bigg)+\alpha c-r(Y_t)\\
  &+\frac{\alpha}{\alpha-\eta(t,Y_t)}\theta( Y_t)^*\left(\sigma_f(Y_t)^*D\phi(t,Y_t)+\frac{1}{\alpha} \eta(t,Y_t)\theta(Y_t)\right)+\frac{1}{2}|\theta(Y_t)|^2\\
  &-\lambda \int_{z>0}\left({\mathrm e}^{\{\alpha-\eta(t,Y_t)\} z  }-1\right) \nu(dz)\bigg\}dt+\tilde q_t^*dW_t+\int_{z>0} h_{t}(z)  N(dt,dz),\\
\end{split}
\end{align}
and 
\begin{align}
\begin{split}
  &d\tilde p_t=\bigg[ X^{\hat\pi}_t\bigg\{\frac{\partial \eta}{\partial t}(t,Y_t)+\frac{1}{2}{\rm tr}(\sigma_f(Y_t)\sigma_f(Y_t)^*D^2\eta(t,Y_t))+g(Y_t)^*D\eta(t,Y_t)\\
 &+r(Y_t)\eta(t,Y_t)+\frac{\eta(t,Y_t)}{\alpha-\eta(t,Y_t)}\bigg(\theta(Y_t)+\sigma_f(Y_t)^* \frac{D\eta(t,Y_t)}{\eta(t,Y_t)}\bigg)^*  \sigma_f(Y_t)D\eta(t,Y_t) \bigg\}\\
  &+\frac{\partial \phi}{\partial t}(t,Y_t)+\frac{1}{2}{\rm tr}(\sigma_f(Y_t)\sigma_f(Y_t)^*D^2\phi(t,Y_t))+g(Y_t)^*D\phi(t,Y_t)+c\eta(t,Y_t)\\
  &+\frac{\eta(t,Y_t)}{\alpha-\eta(t,Y_t)}\bigg(\theta(Y_t)+\sigma_f(Y_t)^* \frac{D\eta(t,Y_t)}{\eta(t,Y_t)}\bigg)^*\bigg(\theta(Y_t)+\sigma_f(Y_t)^*D\phi(t,Y_t)\bigg) \bigg] dt\\
 &+ \tilde q_t^*dW_t+\int_{z>0} h_{t}(z)  N(dt,dz).\\
\label{BSDE-multi-4} 
\end{split}
\end{align}
By comparing the coefficients of the first order and the zero order terms of $X^{\hat\pi}_t$ in \eqref{BSDE-multi-3} and \eqref{BSDE-multi-4}, we see that $\eta(t,Y_t)$ and $\phi(t,Y_t)$ must satisfy $(\ref{PDE-FBSDE-eta-sec})$ and $(\ref{PDE-FBSDE-phi-sec} )$.

Finally, we show the solution of the optimal trajectory. The ansatz for $p_t$ and $\tilde X^{\hat\pi}_t=-{\mathrm e}^{-\alpha X^{\hat\pi}_t}$ imply 
\be\label{Xp-1}
\log(-\tilde X^{\hat\pi}_tp_t)= -(\alpha-\eta(t,Y_t)) X^{\hat\pi}_t+\phi(t,Y_t).
\en
By using \eqref{FSDE-multi} and \eqref{BSDE-multi}, we have 
\begin{align}\label{Xp-2}
\begin{split}
\log(-\tilde X^{\hat\pi}_tp_t)&=\log(-\tilde X^{\hat\pi}_0p_0 )-\int_0^tr(Y_s)ds-\int_0^t\theta(Y_s)^*dW_s\\
 &-\int_0^t\frac{1}{2}\left|\theta(Y_s)\right|^2ds-\int_0^t\int_{z>0}\{\alpha-\eta(s-, Y_{s-}) \}N(ds,dz)\\
&-\lambda \int_0^t \int_{z>0} \left({\mathrm e}^{\{\alpha-\eta(s-,Y_{s-}) \}z} -1 \right)\nu(dz)ds.
\end{split}
\end{align}
Identifying \eqref{Xp-1} and \eqref{Xp-2} gives the optimal trajectory written as $\eqref{opX}$.

\section{The Proof of Theorem \ref{uniqueness}}   \label{proof-uniqueness}
We prove the uniqueness of the solution of FBSDEs $(\ref{FSDE-multi})$-$(\ref{BSDE-multi})$.
Using $(\ref{FSDE-multi})$ and $(\ref{BSDE-multi})$, we have
\begin{align*}
d&(\tilde{p}_t-\alpha X_t^{\hat{\pi}} )=-\left\{r(Y_t)+\frac{1}{2} |\theta(Y_t)|^2+\lambda \int_{z>0} \{ {\mathrm e}^{\alpha z+h_t(z)}-1\}\nu(dz) \right\}dt\\
&-\theta(Y_t)^*dW_t+\int_{z>0} \{h_t(z)+\alpha z\}N(dt, dz).
\end{align*}
Therefore, we have
\begin{align*}
U(X^{\hat{\pi}}_T)=p_T U(X^{\hat{\pi}}_T)=-{\mathrm e}^{\tilde{p}_T-\alpha X^{\hat{\pi}}_T }
=-{\mathrm e}^{\tilde{p}_0-\alpha x }{\mathrm e}^{-\int^T_0r(Y_t)dt}\ \eta^h_T,
\end{align*}
where $\eta_t^h$ is defined by
\begin{align}\label{eta}
\eta_t^h:=&{\mathrm e}^{-\int^t_0 \theta(Y_s)^*dW_s-\frac{1}{2}\int^t_0|\theta(Y_s)|^2ds} \\
&\cdot {\mathrm e}^{\int^t_0\int_{z>0}(h_s(z)+\alpha z)N(ds,dz)-\lambda \int^t_0\int_{z>0}({\mathrm e}^{h_s(z)+\alpha z}-1 )\nu(dz) ds  }. \nonumber
\end{align}
Here, using $\eqref{htz-condition}$ and the arguments of proving $\eqref{tildepiA}$, we see that $E\left[ \eta_t^h\right]=1$. Hence, we can define the probability measure $P^h$ by
\begin{align*}
\frac{dP^h}{dP}\bigg |_{{\mathcal{F}}_T}&=\eta_T^h,
\end{align*}
Then, we have
\begin{align*}
E\left[ U(X^{\hat{\pi}}_T)\right]={\mathrm e}^{\tilde{p}_0}U(x) E^{h}\left[ {\mathrm e}^{-\int^T_0r(Y_t)dt} \right],
\end{align*}
where $E^h[\cdot]$ is the expectation of the probability measure $P^h$.

Prepare another solution $(X'_t, \tilde{p}_t', \tilde{q}_t', h_t'(z)) \in {\mathbb S}^{2} \times {\mathbb S}^{2} \times {\mathbb L}^{2} \times {\mathbb L}_N^{2}$ of FBSDEs $(\ref{FSDE-multi})$ and $(\ref{BSDE-multi})$ : 
\begin{align}
 d&X_t'=\bigg\{r(Y_t)X_t'+c+\frac{1}{\alpha}\bigg(|\theta(Y_t)|^2+ \theta(Y_t)^*\tilde q_t'\bigg)\bigg\}dt \label{FSDE-multi'}\\
 & +\frac{1}{\alpha}\left(\theta(Y_t)+\tilde q_t'\right)^* dW_t-\int_{z>0} z   N(dt,dz), \nonumber \\
 d& \tilde p_t'=\bigg\{\alpha c+r(Y_t)(\alpha X_t'-1)+\frac{1}{2}|\theta(Y_t)|^2+\theta(Y_t)^*\tilde q_t'\bigg\}dt \label{BSDE-multi'}\\
 & -\lambda \int_{z>0} \{ {\mathrm e}^{\alpha z+h_t'(z)}-1\}\nu(dz)dt +(\tilde q_t')^*dW_t+\int_{z>0}h_{t}'(z)  N(dt,dz). \nonumber
\end{align}
From $(\ref{BSDE-multi})$ and $(\ref{FSDE-multi'})$, we have
\begin{align}\label{PX'}
d(\tilde{p}_t-\alpha X'_t)&=d(\tilde{p}_t-\alpha \color{blue}X^{ \hat\pi}_t\color{black})+\alpha d(X^{\hat{\pi}}_t-X_t') \\
&=-r(Y_t)dt+d\log \eta^h_t+\alpha  d(X^{\hat{\pi}}_t-X_t'). \nonumber 
\end{align}
Further, we have
\begin{align*}
\alpha  d(X^{\hat{\pi}}_t-X_t')=\alpha r(Y_t) (X^{\hat{\pi}}_t-X_t')dt+(\tilde{q}_t-\tilde{q}_t')^*dW^N_t,
\end{align*}
where $W^N_t$ is given by $(\ref{hatNW})$ and is a Brownian motion under $\eqref{hatPN}$ in and also $P^h$ given here.
Hence, we have 
\begin{align}\label{MB}
{\mathrm e}^{-\int_0^T r(Y_t) dt} \alpha (X^{\hat{\pi}}_T-X'_T)=\int_0^T {\mathrm e}^{-\int_0^t r(Y_s) ds} (\tilde{q}_t-\tilde{q}_t')^*dW^N_t.
\end{align}
Using this and Jensen's inequality, we have
\begin{align}
&\frac 1{E^h \left[{\mathrm e}^{-\int_0^T r(Y_t) dt}\right] } E^h \left[ {\mathrm e}^{-\int_0^T r(Y_t) dt} {\mathrm e}^{\alpha (X^{\hat{\pi}}_T-X'_T} ) \right]  \nonumber \\
\geq & \exp\left( \frac 1{E^h \left[{\mathrm e}^{-\int_0^T r(Y_t) dt} \right] } E^h\left[ {\mathrm e}^{-\int_0^T r(Y_t) dt} \alpha(X^{\hat{\pi}}_T-X'_T)\right] \right) \label{Ji} \\
= & \exp\left( \frac 1{E^h \left[{\mathrm e}^{-\int_0^T r(Y_t) dt} \right] } E^h\left[ \int_0^T {\mathrm e}^{-\int_0^t r(Y_s) ds} (\tilde{q}_t-\tilde{q}_t')^*dW^N_t \right] \right)  \nonumber \\
=&1. \nonumber
\end{align}
On the other hand, by $\eqref{PX'}$, we have
\begin{align*}
U(X_T')=-{\mathrm e}^{\tilde{p}_0-\alpha x }{\mathrm e}^{-\int^T_0r(Y_t)dt}\ \eta^h_T {\mathrm e}^{\alpha(X^{\hat{\pi}}_T-X_T')},
\end{align*}
and
\begin{align*}
E\left[U(X_T') \right]={\mathrm e}^{\tilde{p}_0}U(x)E^h\left[ {\mathrm e}^{-\int^T_0r(Y_t)dt} {\mathrm e}^{\alpha(X^{\hat{\pi}}-X_t')} \right].
\end{align*}
Therefore, we have
\begin{align*}
E\left[U(X_T') \right]\le E\left[ U(X^{\hat{\pi}}_T)\right].
\end{align*}
Here, we use $\eqref{FSDE-multi'}$ and $\eqref{Ji}$. Furthermore, we apply the above arguments to
\begin{align*}
d(\tilde{p}'_t-\alpha X^{\hat{\pi}}_t),
\end{align*}
and
\begin{align*}
d(\tilde{p}'_t-\alpha X^{\hat{\pi}}_t)&=d(\tilde{p}'_t-\alpha X'_t)+\alpha d(X'_t-X^{\hat{\pi}}_t).
\end{align*}
Hence, we have
\begin{align*}
E\left[U(X_T') \right]\ge E\left[ U(X^{\hat{\pi}}_T)\right].
\end{align*}
As a result, we have
\begin{align*}
E\left[ U(X^{\hat{\pi}}_T)\right]=E\left[U(X_T') \right].
\end{align*}
This is equivalent to the relation
\begin{align}
&\frac 1{E^h \left[{\mathrm e}^{-\int_0^T r(Y_t) dt}\right] } E^h \left[ {\mathrm e}^{-\int_0^T r(Y_t) dt} {\mathrm e}^{\alpha (X^{\hat{\pi}}_T-X'_T} ) \right]  \nonumber \\
= & \exp\left( \frac 1{E^h \left[{\mathrm e}^{-\int_0^T r(Y_t) dt} \right] } E^h\left[ {\mathrm e}^{-\int_0^T r(Y_t) dt} \alpha(X^{\hat{\pi}}_T-X'_T)\right] \right)   \nonumber  \\
=&1. \nonumber
\end{align}
Hence, $X^{\hat{\pi}}_T-X'_T=0, a.s.$. Using $\eqref{MB}$, we see that $\tilde{q}_t-\tilde{q}'_t=0, a.s.$. And, we have
\begin{align*}
0=E^h\left[{\mathrm e}^{-\int_0^T r(Y_t) dt} (X^{\hat{\pi}}_T-X'_T) \Bigl| \mathcal{F}_t \right]={\mathrm e}^{-\int_0^t r(Y_s) ds} (X^{\hat{\pi}}_t-X'_t),
\end{align*}
which leads to
\begin{align*}
X^{\hat{\pi}}_t-X'_t=0, a.s..
\end{align*}
Here we use the property that the martingale property from $\eqref{MB}$. Now, we compare $\eqref{BSDE-multi}$ with $\eqref{BSDE-multi'}$.
Recall
\begin{align}\label{JJ}
\begin{split}
&{\mathrm e}^{\tilde{p}_t-\alpha X^{\hat{\pi}}_t+\int^t_0r(Y_s)ds}={\mathrm e}^{\tilde{p}_0-\alpha x} \eta^h_t, \\
&{\mathrm e}^{\tilde{p}'_t-\alpha X'_t+\int^t_0r(Y_s)ds}={\mathrm e}^{\tilde{p}_0-\alpha x} \eta^{h'}_t,
\end{split}
\end{align}
where $\eta^{h'}_t$ is given in $\eqref{eta}$ where $h$ is replaced by $h'$. 
Recalling that $X^{\hat{\pi}}_t=X'_t$ and $\tilde{p}_T=\tilde{p}'_T$, we see that $\eta^h_T=\eta^{h'}_T$.
Since $\eta^h_t$ and $\eta^{h'}_t$ are martingale, we have
\begin{align}\label{etamarteq}
\eta^h_t=E[\eta^h_T|\mathcal{F}_t]=E[\eta^{h'}_T|\mathcal{F}_t]=\eta^{h'}_t.
\end{align}
From $\eqref{JJ}$ we have
\begin{align*}
\tilde{p}_t=\tilde{p}'_t, \quad 0\le t \le T.
\end{align*}
Moreover, we have
\begin{align*}
\eta^h_T&=1-\int^T_0\eta^h_t \theta(Y_t)^*dW_t+\int^T_0 \eta^h_{t-}\int_{z>0} \{h_{t-}(z)+\alpha z \}\tilde{N}(dt,dz), \\
\eta^{h'}_T&=1-\int^T_0\eta^{h'}_t \theta(Y_t)^*dW_t+\int^T_0 \eta^{h'}_{t-}\int_{z>0} \{h'_{t-}(z)+\alpha z \}\tilde{N}(dt,dz). 
\end{align*}
Using $\eqref{etamarteq}$, we have
$$\int^T_0 \eta^h_{t-}\int_{z>0} \{h_{t-}(z)- h'_{t-}(z)\}\tilde{N}(dt,dz)=0,$$
which leads to $h_{t}(z)- h'_{t}(z)=0, a.s. dt \times \nu(dz)$.

\section{The Proof of Theorem \ref{Thm-verification-FBSDE}}\label{proof-verification-FBSDE}

We verify $\hat{\pi}_t\in \tilde{\mathcal{A}}_T$. Under $\hat{P}^N$, $\tilde{N}^N(t, E), E \in \mathcal{B}((0, \infty))$ defined by
$$\tilde{N}^N(t, E):=\tilde{N}(t, E)-\lambda \int^t_0 \int_{E} \left(1-{\mathrm e}^{\{\alpha-\eta(s,y) \}z} \right)\nu(dz)ds$$
is a $\hat{P}^N$-martingale. 
Under $\hat{P}^N$ we recall that
\begin{align*}
dX^{\hat{\pi}}_t=&r(Y_t)X^{\hat{\pi}}_t dt+cdt+\hat{\pi}^*_t\sigma_p(Y_t)d\hat{W}^N_t-\int_{z>0}z \tilde{N}^N(dt,dz) \\
&-\lambda\int_{z>0}z\left(2-{\mathrm e}^{\{\alpha-\eta(t,y)\}z}  \right)\nu(dz)dt.
\end{align*}
And, we get 
\begin{align*}
&X^{\hat{\pi}}_T= {\mathrm e}^{\int^T_0 r(Y_t)dt}x+\int^T_0 {\mathrm e}^{\int^T_t r(Y_s)ds}cdt+\int^T_0 {\mathrm e}^{\int^T_t r(Y_s)ds}\hat{\pi}^*_t \sigma_p(Y_t)d\hat{W}^N_t \\ 
&-\int^T_0 {\mathrm e}^{\int^T_t r(Y_s)ds} \int_{z>0} z\tilde{N}^N(dt,dz)-\lambda \int^T_0 {\mathrm e}^{\int^T_t r(Y_s)ds} \int_{z>0} z \left(2-{\mathrm e}^{\{\alpha-\eta(t,y)\}z}  \right)\nu(dz)dt .
\end{align*}
Hence, 
we have
\begin{align*}
\hat{E}^N&\left[\int^T_0 \hat{\pi}^*_t\sigma_p(Y_t)\sigma_p(Y_t)^* \hat{\pi}_t dt  \right] \le K \bigg[ x^2+1+ \hat{E}^N[(X_T^{\hat{\pi}})^2] + \int^T_0 \int_{z>0} z^2 \nu(dz) dt \bigg].
\end{align*}
From $\eqref{opX}$ we recall, under $\hat{P}^N$
\begin{align}
\begin{split}
& X_t=X^{\hat{\pi}}_t=\frac{1}{\alpha-\eta(t,Y_t)}\left((\alpha-\eta(0,y))x+\phi(t,Y_t)-\phi(0,y)\right)\\
 &+\frac{1}{\alpha-\eta(t,Y_t)}\bigg(\int_0^tr(Y_s)ds+\int_0^t\theta(Y_s)^*d\hat{W}^N_s-\int_0^t\frac{1}{2}\left|\theta(Y_s)\right|^2ds\\
 &-\int_0^t\int_{z>0} \{\alpha-\eta(s, Y_{s}) \} z \tilde{N}^N(dt,dz) \\
 &-\int_0^t\int_{z>0} \{\alpha-\eta(s, Y_{s}) \}  \left(2-{\mathrm e}^{\{\alpha-\eta(s,Y_s)\}z}  \right)\nu(dz)ds  \\
&+\lambda \int_0^t \int_{z>0}({\mathrm e}^{-\{\alpha-\eta(s, Y_{s})\} z}-1)\nu(dz)ds \bigg).
\end{split}
\end{align}
From $(\ref{phisol})$ we have$|\phi(t,y)|\le K(1+|y|^2)$.
Then, we can see that $\hat{E}^N[(X_T^{\hat{\pi}})^2] <\infty$.
Therefore, we have
\begin{align*}
\hat{E}^N&\left[\int^T_0 \hat{\pi}^*_t\sigma_p(Y_t)\sigma_p(Y_t)^* \hat{\pi}_t dt  \right] < \infty,
\end{align*}
which leads to $\hat{\pi}_t\in \tilde{\mathcal{A}}_T$.


\begin{thebibliography}{99}
\bibitem{AM2009} Azcue,P., Muler,N. :
Optimal investment strategy to minimize the ruin probability of an insurance
company under borrowing constraints.
Insurance Math. Econom. {\bf 44}(1), 26--34 (2009)

\bibitem{KMY2017}
Bahlali, K., Eddahbi, M., Ouknine, Y. :
  Quadratic {BSDE} with {$L^2$}-terminal data: Krylov's estimate,
  {I}t\^o-{K}rylov's formula and existence results.
 Ann. Probab, {\bf 45}(4), 2377--2397 (2017)
 
\bibitem{PE2013}
Barrieu, P., El Karoui, N. :
Monotone stability of quadratic semimartingales with applications to
unbounded general quadratic BSDEs.
Ann. Probab. {\bf 41}(3B), 1831--1863 (2013)

\bibitem{BG2008} Bai,L., Guo,J. :
Optimal proportional reinsurance and investment with multiple risky assets
and no-shorting constraint.
Insurance Math. Econom. {\bf 42}(3), 968--975 (2008)

\bibitem{BHLT2014} Belkina,T., Hipp,C., Luo, S., Taksar,M. :
Optimal constrained investment in the Cramer-
Lundberg model.
Scand. Actuar. J. {\bf 5}, 383--404 (2014)

\bibitem{Bens1992} Bensoussan, A. : 
Stochastic Control of Partially Observable Systems. 
Cambridge University Press, Cambridge (1992)

\bibitem{BF2013} Badaoui, M., Fern$\acute{{\rm a}}$ndez, B. :
An optimal investment strategy with maximal risk aversion and its ruin probability in the presence of stochastic volatility on investments.
Insurance Math. Econom. {\bf 53}(1), 1--13 (2013) 

\bibitem{BFS2018} Badaoui, M., Fern$\acute{{\rm a}}$ndez, B., Swishchuk,A. :
An Optimal Investment Strategy for Insurers in Incomplete Markets.
Risks. {\bf 6}(2), 31, (2018)



\bibitem{BH2006}
P.~Briand, P.,Hu, Y. :
{BSDE} with quadratic growth and unbounded terminal value.
 Probab. Theory Relat. Fields. {\bf 136}, 604--618 (2006)

\bibitem{BH2008}
P.~Briand, P., Hu, Y. :.
Quadratic {BSDE}s with convex generators and unbounded terminal
 conditions.
 Probab. Theory Relat. Fields. {\bf 141}, 543--567 (2008)
 
 



\bibitem{BW2017} Bo, L., Wang, S. :
Optimal investment and risk control for an insurer with stochastic
factor.
Oper. Res. Lett. {\bf 45}(3), 259--265 (2017) 



\bibitem{B1995} Browne S. 
Optimal investment policies for a firm with a random risk process: Exponential utility and minimizing the probability of ruin ,
Math. Oper. Res. {\bf 20}(4), 937--958 (1995)


\bibitem{CH2011} Cheridito, P., Hu. Y.: 
Optimal consumption and investment in incomplete markets with general
  constraints.
Stochastics and Dynamics {\bf 11}(2,3), 283--299 (2011)

\bibitem{CN2014}
Cheridito, P., Nam, K. :
{BSDE}s with terminal conditions that have bounded Malliavin derivative.
J. Funct. Anal. {\bf 266}(3), 1257--1285 (2014)

\bibitem{CN2015}
Cheridito, P., Nam, K. :
Multidimensional quadratic and subquadratic {bsde}s with special
  structure.
 Stochastics 87(5), 871--884 (2015)

\bibitem{De2002}
Delarue, F. :
On the existence and uniqueness of solutions to
FBSDEs in a non-degenerate case.
Stoch. Proc. Appl.. {\bf 99}, 209--286 (2002)

\bibitem{DHB2011}
Delbaen, F., Hu, Y., Bao, X. :
Backward {SDE}s with Superquadratic Growth.
Probab. Theory Relat. Fields {\bf 150}, 145--192 (2011)



 
\bibitem{FHMS2008} Fern$\acute{{\rm a}}$ndez, B., Hern$\acute{{\rm a}}$ndez, D., Meda, A., Saavedra, P. :
An optimal investment strategy with maximal risk aversion and its ruin probability.
Math. Meth. Oper. Res. {\bf 68}(1), 159--179 (2008)

\bibitem{FS2006} Fleming, W.H., Soner, M. :
Control Markov Processes and Viscosity Solutions, 2nd Edition.
Springer-Verlag, New York (2006)

\bibitem{Frei2014}
Frei, C. :
Splitting Multidimensional BSDEs and finding local equilibria.
Stoch. Proc. Appl. {\bf124}(8),
2654–2671 (2014)

\bibitem{GL2014} Guan, G., Liang, Z. :
Optimal reinsurance and investment strategies for insurer under
interest rate and inflation risks.
Insurance Math Econom. {\bf 55}(2), 105--115. 


\bibitem{HNS2017} Hata, H., Nagai, H., Sheu, S.J. :
An optimal consumption problem for general factor models.
SIAM J. Cont. Optim. {\bf 56}(5), 3149--3183 (2018)


\bibitem{HY2017} Hata, H., Yasuda, K. :
Expected exponential utility maximization of insurers with a linear Gaussian stochastic factor model. Scand. Actuar. J. {\bf 2018}(5), 357--378 (2018)

\bibitem{HKT2016} Heyne, G., Kupper, M., Tangpi, L. :
Portfolio optimization under nonlinear utility.
Int. J. Theor. Appl. Finance {\bf 19}(5), 283--299 (2016) 

\bibitem{HP2000} Hipp, C., Plum, M. :
Optimal investment for insurers.
Insurance Math Econom. {\bf 27}(2), 215--228 (2000) 

\bibitem{HS2004} Hipp, C., Schmidli, H. : 
Asymptotics of ruin probabilities for controlled risk processes in the small
claims case,
Scand Actuar J. {\bf 5}, 321--335 (2004).


\bibitem{HI2014} Horst, U., Imkeller, P., R \'eveillac, Zhang, J. :
Forward \& backward systems for expected utility maximization.
Stochastic Process. Appl. {\bf 124}, 1813--1848 (2014)

\bibitem{HU2005} Hu, Y., Imkeller, P., M\"uller, M :
Utility maximization in incomplete markets. 
Ann. Appl. Probab. {\bf 15}(3), 1691--1712 (2005)





\bibitem{YS2016}
Hu, Y., Tang, S. :
 Multi-dimensional backward stochastic differential equations of
  diagonally quadratic generators.
 Stoch. Proc. Appl. {\bf 126}(4), 1066--1086 (2016)


\bibitem{HYZ2016} Huang, Y., Yang, X., Zhou, J. :
Optimal investment and proportional reinsurance for a
jump-diffusion risk model with constrained control variables,
J. Comput. Appl. Math. {\bf 296}, 443--461 (2016)

\bibitem{IR2010}
Imkeller, P.,  Reis, G.~D. :
Path regularity and explicit convergence rate for {BSDE} with
  truncated quadratic growth.
 Stoch. Proc. Appl. {\bf 120}, 348--379 (2010)
 
 \bibitem{JKP2014}
 Jammneshan, A., Kupper, M., Luo, P. :
 Multidimensional quadratic BSDEs with separated generators.
 Preprint (2014) 

\bibitem{Karatzas2000}
 Karatzas, I., Shreve, S. E. :
Brownian Motion and Stochastic Calculus Second Edition.
Springer-Verlag, New York, (1998) 

\bibitem{Ko2000}
Kobylanski, M. :
Backward stochastic differential equations and partial differential
  equations with quadratic growth,
Ann. Probab. {\bf 28}(2), 558--602 (2000)
 
 

\bibitem{K1980}
Krylov, N.Y. : 
Controlled Diffusion Processes.
Springer-Verlag, New York, Heidelberg, Berlin, (1980) 

 


\bibitem{KLT2017}
Kupper, M., Luo, P., Tangpi, L. :
Multidimensional Markovian FBSDEs with super- quadratic growth.
Preprint (2017)





\bibitem{LRZ2015} Li, D., Rong, X., Zhao, H. :
Time-consistent reinsurance-investment strategy for a
mean-variance insurer under stochastic interest rate model and
inflation risk.
Insurance Math Econom. {\bf 64}, 28--44 (2015) 

\bibitem{LL2013} Li, T., Li, Z. :
Optimal time-consistent investment and reinsurance strategies for
mean–variance insurers with state dependent risk aversion.
Insurance Math Econom. {\bf 53}(1), 86--97 (2013)

\bibitem{LZL2012} Li, Z., Zeng, Y., Lai, Y. :
Optimal time-consistent investment and reinsurance strategies for insurers
under Heston's SV model.
Insurance Math Econom. {\bf 51}(1), 191--203 (2012)

\bibitem{LB2014} Liang, Z., Bayraktar, E. :
Optimal reinsurance and investment with unobservable claim size
and intensity.
Insurance Math Econom. {\bf 55}, 156--166 (2014)

\bibitem{LYG2011} Liang.Z., Yuen, K.C., Guo, J. : 
Optimal proportional reinsurance and investment in a stock market with
Ornstein–Uhlenbeck process. 
Insurance Math Econom. {\bf 49}(2), 207--215 (2011)

\bibitem{LT2015}
Luo, P., Tangpi, L. : 
Solvability of coupled FBSDEs with quadratic and
superquadratic growth.
arXiv: 1505.01796v1 (2015)

\bibitem{LT2017}
Luo, P., Tangpi, L. :
Solvability of coupled FBSDEs with diagonally
quadratic generators.
Stochastics and Dynamics. {\bf 17}(6), (2017)



\bibitem{L2008} Luo,S. :
Ruin minimization for insurers with borrowing
constraints.
N. Am. Actuar. J. {\bf 12}(2), 143--174 (2008) 

\bibitem{LTT2008} Luo,S., Taksar, M., Tsoi.A. :
On reinsurance and investment for large insurance portfolios.
Insurance Math Econom. {\bf 42}(1), 434--444 (2008). 

\bibitem{LWZ2016}Luo,S., Wang,M., Zeng, X. :
Optimal reinsurance: minimize the expected time to reach a
goal.
Scand. Actuar. J. {\bf 8}, 741--762 (2016).

\bibitem{MWZZ2015}
Ma, J., Wu, Z., Zhang, D., Zhang, J. :
ON WELL-POSEDNESS OF FORWARD–BACKWARD SDES—A UNIFIED APPROACH. 
Ann. Appl. Probab. {\bf 25}(4), 2168--2214 (2015)

\bibitem{MM2010}
Morlais, M.-A. :
A new existence result for quadratic BSDEs with jumps with application to the utility maximization problem. 
Stoch. Proc. Appl. {\bf 120}, 1966--1995 (2010)



\bibitem{N2003} Nagai, H. :
Optimal strategies for risk-sensitive portfolio optimization problems for
general factor models.
SIAM J. Cont. Optim. {\bf 41}(6), 1779--1800 (2003)

\bibitem{Pham2009}
Pham, H. : 
Continuous-time Stochastic Control and Optimization with
  Financial Applications. 
Springer-Verlag Berlin Heidelberg, (2012) 

\bibitem{S2001} Schmidli, H. : 
Optimal proportional reinsurance policies in a
dynamic setting.
Scand. Actuar. J. {\bf 22}(1), 55--66 (2001)

\bibitem{S2002} Schmidli, H. : 
On minimizing the ruin probability by investment and reinsurance..
Ann. Appl. Probab.  {\bf 12}(2), 890--907 (2002)

\bibitem{Sekine2006} Sekine, J. :
On exponential hedging and related backward stochastic differential
  equations.
Applied Math. \& Optim. {\bf 54}, 131--158 (2003)

\bibitem{PP1990}
Pardoux, \'E., Peng, S. G. :
Adapted Solution of a Backward Stochastic Differential Equation.
Systems Control Lett. {\bf 14}(1), 55--61 (1990)

\bibitem{PW1999}
Peng, S., Wu, Z. :
Fully coupled forward–backward stochastic differential equations and applications
to optimal control.
SIAM J. Control Optim. {\bf 37}(3), 825--843 (1999)

\bibitem{SZ2015} Shen,Y., Zeng, Y. :
Optimal investment-reinsurance strategy for mean-variance insurers
with square-root factor process.
Insurance Math Econom. {\bf 62}(2), 118--137  (2015)

\bibitem{SRZ2014} Sheng D.L., Rong, X., Zhao, H. :
Optimal Control of Investment-Reinsurance
Problem for an Insurer with Jump-Diffusion Risk Process:
Independence of Brownian Motions.
Abstr. Appl. Anal. 2014, Art. ID 194962, 19 pp. (2014)

\bibitem{TE2008}
Tevzadze, R. :
Solvability of backward stochastic differential equations with
 quadratic growth.
 Stochastic Processes and their Applications. {\bf 118}(3),503--515 (2008)


\bibitem{TM2003} Taksar, M., Markussen, C. :
Optimal dynamic reinsurance policies
for large insurance portfolios.
Finance Stochast. {\bf 7}(1), 97--121 (2003)

\bibitem{Touzi2012}
Touzi, N. :  
Optimal stochastic control, stochastic target problems, and
  backward SDE. Springer-Verlag New York. (2012) 


\bibitem{W2007} Wang, N. : 
Optimal investment for an insurer with exponential utility preference.
Insurance Math
Econom. {\bf 40}(1), 77--84 (2007) 

\bibitem{WRZ2018} Wang,Y., Rong, X., Zhao,H. :
Optimal investment strategies for an insurer and a reinsurer
with a jump diffusion risk process under the CEV model
J. Comput. Appl. Math. {\bf 328}, 414--431 (2018) 

\bibitem{WXZ2007} Wang. Z., Xia, J., Zhang. L. : 
Optimal investment for an insurer: The martingale approach. Insurance Math
Econom. {\bf 40}(2), 322--334 (2007) 



\bibitem{XZY2017} Xu, L, Zhang, L., Yao, D. : 
Optimal investment and reinsurance for an insurer under
Markov-modulated financial market.
Insurance Math
Econom. {\bf 74}(1), 7--19 (2017) 

\bibitem{YZ2005} Yang, H., Zhang, L.:
Optimal investment for insurer with jump-diffusion risk process.
Insurance Math Econom. {\bf 37}(3), 615--634 (2005) 


\bibitem{ZMZ2016} Zhang, X., Meng, H., Zeng, Y. :
Optimal investment and reinsurance strategies for insurers with
generalized mean-variance premium principle and no-short selling.
Insurance Math Econom. {\bf 67}(3), 125--132 (2016)

\bibitem{ZRZ2013} Zhao, H., Rong, X., Zhao, Y. :
Optimal excess-of-loss reinsurance and investment problem for an
insurer with jump-diffusion risk process under the Heston model.
Insurance Math Econom. {\bf 53}(3), 504--514 (2013)

\bibitem{ZL2011} Zeng, Y., Li, Z. :
Optimal time-consistent investment and reinsurance policies for mean-variance
insurers.
Insurance Math Econom. {\bf 49}(1), 145--154 (2011)

\bibitem{ZC2014} Zou,B., Cadenillas, A. :
Optimal investment and risk control policies for an insurer:
Expected utility maximization.
Insurance Math Econom. {\bf 58}(3), 57--67 (2014)



\bibitem{Z2009} Zhou, Q. : 
Optimal investment for an insurer in the L$\acute{{\rm e}}$vy market: The martingale
approach.
Stat. Probab. Lett. {\bf 79}(14), 1602--1607 (2009) 

\end{thebibliography}
\end{document}